\documentclass[preprint,12pt,authoryear]{elsarticle}

\usepackage{amssymb}

\journal{Journal of Forecasting}

\begin{document}

\begin{frontmatter}

\title{Backward-in-Time Selection of the Order of Dynamic Regression Prediction Model}

\author[asu]{Ioannis˜Vlachos}
\ead{ivlachos@asu.edu}
\author[auth]{Dimitris Kugiumtzis\corref{cor1}}
\ead{dkugiu@gen.auth.gr}
\cortext[cor1]{Corresponding author}
\address[asu]{School Biological \& Health Systems Engineering,
Ira Fulton School of Engineering, Arizona State University, Tempe, Arizona, USA}
\address[auth]{Department of Mathematical, Physical and Computational Sciences of Engineering, Aristotle University of Thessaloniki, Greece}


\begin{abstract}
We investigate the optimal structure of dynamic regression models used in multivariate time series prediction and propose a scheme to form the lagged variable structure called Backward-in-Time Selection (BTS) that takes into account feedback and multi-collinearity, often present in multivariate time series. We compare BTS to other known methods, also in conjunction with regularization techniques used for the estimation of model parameters, namely principal components,
partial least squares and ridge regression estimation. The predictive efficiency of the different models is assessed by means of Monte Carlo simulations for different settings of feedback and multi-collinearity. The results show that BTS has consistently good prediction performance while other popular methods have varying and often inferior performance. The prediction performance of BTS was also found the best when tested on human electroencephalograms of an epileptic seizure, and to the prediction of returns of indices of world financial markets.
\end{abstract}


\end{frontmatter}

\section{Introduction\label{intro}}
For the prediction of a multivariate time series it is important to model the
interactions and lag dependencies that exist between the different variables of the
system. These variables can reflect a single quantity measured at different
``locations'', e.g. channels of an electroencephalogram (EEG) recording \citep{Prado01} or different but
connected quantities, e.g. financial products \citep{Cramer76,Welch84,Bentzen01} or
physiological indices of a patient \citep{Rigney93,Johnston94}.

A straightforward approach in the analysis of multivariate time series is to extend
models and methods used in univariate time series analysis, e.g. the autoregressive models for univariate time series to the vector autoregressive (VAR) models and the dynamic regression (DR) models for multivariate time series \citep{Pankratz91,Luetkepohl05,Wei06}. The extension, though seemingly straightforward, entails some caution
with regard to data conditions that are likely to be present in multivariate time series. Such conditions are
bidirectional or unidirectional feedback between the time series, i.e. a variable depends
on another variable at a delayed time, which in turn may depend on the first variable at
a further delayed time \citep{Hsiao82}, and multi-collinearity, i.e. strong linear
relationships between the time series \citep{Salinas87,VanDenPoel04}. Both these
conditions create spurious correlations between the different components of the
multivariate time series and spurious autocorrelations on the single time series, which
can result to misspecified models \citep{Sapsford06}.

Many of the methods for model selection for multivariate time series stem from multiple regression \citep{Leamer78,Castle11}, such as criteria trading-off fitting accuracy to model complexity, e.g. Akaike's information criterion and final prediction error \citep[chap. 4]{Luetkepohl05}, shrinkage (or regularization) techniques such as principal component regression (PCR) \citep{Jolliffe82}, stepwise regression \citep{Hocking76} and Bayesian model averaging \citep{Hoeting99}. However, feedback and multi-collinearity requires a suitable adaptation of these criteria in order to identify the most appropriate set of lags from each time series, referred to as subset VAR models. For example, the subset selection schemes using a top-down or bottom-up strategy are in analogue to forward and backward selection in regression \citep[chap. 5.2.8]{Luetkepohl05}. All the schemes are suboptimal in the sense that they do not explore all possible lag combinations of the variables, and they are mostly based on the idea of optimizing the lag structure for each variable. Here, we propose a scheme of the bottom-up type accounting for feedback and multi-collinearity, but also an inherent property of time series that the dependence structure is closely related to the temporal order of the variables, i.e. temporally close variables are more likely to be correlated than variables falling further apart in time. Thus our scheme, called Backward-in-Time Selection (BTS), evaluates progressively the inclusion of the lagged variables in the model, starting with the most current variables.

We compare the proposed BTS scheme to other methods for model structure. Given that the model structure can be indirectly determined through model estimation, e.g. PCR regularization may give a model structure of reduced degrees of freedom, we include in the study different methods for estimation of dynamic regression models. In particular, we consider ordinary least square estimation (OLS) of the dynamic regression model and the regularization techniques of PCR, partial least squares regression (PLS) and ridge regression (RR) that are commonly used in problems of regression with the aforementioned problems. For all the combinations of schemes for model structure and model estimation, we compare their predictive efficiency with the use of Monte Carlo simulations.

In Section \ref{mulmod}, the basics of linear multivariate modeling are presented. In
Section \ref{BTS}, we present the BTS scheme for determining the model structure and in Section \ref{modmeth} we discuss other methods for model structure as well as methods for model estimation. In Section \ref{mocasi}, Monte Carlo simulations are presented for different processes and the results are discussed. In section \ref{eegdata}, we present applications on two real data sets, one of multi-channel EEG recordings and one of the Morgan
Stanley Capital International (MSCI) indices. Finally, in Section \ref{discus}, overall
conclusions are derived.

\section{Multivariate Modelling\label{mulmod}}

The two most known classes of linear models for multivariate time series are the
\emph{Vector Autoregressive} model (VAR) \citep{Sims80,Wei06} and the \emph{Dynamic
Regression} model (DR) \citep{Pankratz91}.  Both can be seen as extensions of the simple
autoregressive model (AR) for univariate time series, with the distinct difference that
the former class models the vector variable at each time point as a whole, while the latter class
models each component separately. Throughout our analysis we denote the multivariate
series $\{\mathbf{y}_t\}_{t=1}^{N}$ where
$\mathbf{y}_t=\big(y_{1,t},y_{2,t},\ldots,y_{n,t}\big)'$ is the variable vector at time
$t$ comprised of the $n$ components $y_{i,t}$. We focus on one-step-ahead prediction
models but the generalization for $T$-step-ahead is straightforward.

\subsection{Vector Autoregressive Model VAR($k$)\label{varmod}}

The VAR model of order $k$ is the most straightforward modification of the AR model for
multivariate data and is defined as
\begin{equation}
\label{Varmod}
\mathbf{y}_{t+1}=\mathbf{a}_{0}+\mathbf{A}_{1}\mathbf{y}_{t}+\mathbf{A}_{2}\mathbf{y}_{t-1}+\cdots+\mathbf{A}_{k}\mathbf{y}_{t-k+1}
+\mathbf{e}_{t+1},
\end{equation}
where the scalar constant and the scalar coefficients of AR are respectively replaced by
the constant vector $\mathbf{a}_{0}$ and the $n\times n$ coefficient matrices
$\mathbf{A}_{j},\ j=1,\ldots,k$. The vector $\mathbf{e}_t$ has components synchronously
correlated, but time independent with zero mean and finite constant variance. Creating a
matrix containing all the parameters, $\mathbf{A}=\big(\mathbf{a}_{0}\ \mathbf{A}_{1}\
\ldots \ \mathbf{A}_{k}\big)$, and the predictor vector of length $nk+1$,
$\mathbf{x}_t=\big(1,\mathbf{y}_{t}',\mathbf{y}_{t-1}',\ldots,\mathbf{y}_{t-k+1}'\big)'$,
eq.~\ref{Varmod} has the compact matrix form
$\mathbf{y}_{t+1}=\mathbf{A}\mathbf{x}_t+\mathbf{e}_{t+1}$. The Ordinary Least Squares
(OLS) estimation of $\mathbf{A}$ is given by
$$\mathbf{\hat A}=\left(\sum\limits_{t=k}^{N-1}\mathbf{y}_{t+1}{\mathbf{x}_t}'\right)
\left(\sum\limits_{t=k}^{N-1}\mathbf{x}_{t}{\mathbf{x}_t}'\right)^{-1}.$$

The vector form of the VAR model restricts greatly the model estimation, in that all
variables must be included with the same number of lags. Also, with regard to prediction,
an optimum fit for the vector time series may not be optimum for a component time series,
or even for all of them. Another drawback is that since all the parameters are estimated
simultaneously, when their number is large with respect to the number of the available
data, the estimation may become numerically unstable. Workarounds for this problem are
the QR factorization of the matrix with rows
$$[\mathbf{y}_{t+1}',1,\mathbf{y}_{t}',\mathbf{y}_{t-1}',\ldots,\mathbf{y}_{t-k+1}'],$$
or estimating the parameters for each one of the components separately, an approach that
gives consistent estimates \citep{Hamilton94}. When the model in eq.~\ref{Varmod} is
decomposed to scalar equations, studying each one of them separately can also remove the
restriction of having the same lag for each component, which leads to the other type of models, the
Dynamic Regression models.

\subsection{Dynamic Regression Model DR($k_{1},k_{2},\ldots,k_{n}$)\label{drmod}}

The DR model for prediction of one component time series $y_{i,t+1}$ of
$\mathbf{y}_{t+1}$ has the form
\begin{equation}
\label{DRmodel} y_{i,t+1}=a_{i0}+a_{i1}(B)y_{1,t}+\cdots+a_{in}(B)y_{n,t}+e_{t+1}
\end{equation}
where $a_{i0}$ is a constant, $a_{ij}(B)$ for $j=1,2,\ldots,n$ are polynomials of order
$k_{i,j}-1$ with regard to the backshift (or lag) operator $B$ ($By_{i,t}=y_{i,t-1}$) and
$e_t$ are iid. In essence, the prediction of $y_{i,t+1}$ is given by a linear
combination of the components of $\mathbf{y}_{t}$ with different lags for each component.
For the sake of simplicity we will work on centralized time series and omit the constant
parameter.

If all $k_{i,j}$ are equal for all $i$ and $j$ we have the VAR model. Because of this
connection between the VAR and DR models we will study the former as a constrained
version of the latter. Another way to see the DR model is as a type of ordinary multiple
regression where as independent variables we have lagged variables formed by the $n$ time series. Thus
we can apply known methods both from time series analysis and multiple regression for
estimating models of this class.

There are many different expressions for dynamical regression models that are used in
real-world time series analysis. Among them, we note the use of predetermined lag structure
\citep{Welch84}, the inclusion of synchronous values of the other time series to model the
desired one (\citealt{Cramer76,Johnston94}; \citealt[chap. 4]{Luetkepohl05}), the use of transformed variables
\citep{Bentzen01} or the exclusion of lagged values of the modelled time series \citep{Prado01}. All these approaches are based on real-world assumptions about the
data studied. We will work on the general case given by eq.~\ref{DRmodel}.

\section{Backward-in-Time Selection of Model Order \label{BTS}}

The determination of a DR model for a single time series consists of the identification
of appropriate orders $k_{1},k_{2},\ldots,k_{n}$ for the lag polynomials. The proposed scheme
of Backward-in-Time Selection (BTS) aims at selecting the lagged variables progressively and augmenting
the model terms by one lagged variable at a time according to a model selection criterion. The search starts
from the $n$ concurrent variables (lag zero) and goes backward in time towards larger lags until either fitting does not improve or a maximum lag is reached. The rationale with BTS is to select first the terms of the model among the most current variables and then include lagged variables that are least correlated to the lagged variables already selected, and most correlated to the response. Thus when there are several collinear lagged variables, only the variable temporally closer to the response will be selected. In this way, the scheme will result in parsimonious models of small orders as compared to other methods that estimate an order for each variable separately or an order for the VAR model.

The predicted time series is $y_{1,t+1}$, the lags range from 0 up to a maximum lag $K_{\mathrm{max}}$, the model selection criterion used at each step of BTS is the Bayesian Information Criterion (BIC) \citep{Schwarz78} and the estimation of parameters is done by OLS. In detail, BTS determines the appropriate set of orders $k_{1},k_{2},\ldots,k_{n}$ for the $n$ variables as follows. Starting from orders $(0,0,\ldots,0)$ for the $n$ variables we increase by 1 each (we go to $(1,0,\ldots,0)$, $(0,1,\ldots,0)$, etc) and locate the one giving minimum BIC. We select this order vector and repeat the process increasing again by 1 each component and so forth. If for all components the new BIC values are larger than the minimum BIC in the last step we increase each order by 2 in order to encompass cases where intermediate delays have insignificant effect on the response. If there is still no decrease in BIC we increase by 3 etc, until we arrive to $(K_{\mathrm{max}},K_{\mathrm{max}},\ldots,K_{\mathrm{max}})$. If again all new BIC values are larger than the last, we select this last order vector as the vector of optimum model orders.

The algorithm of BTS can be decomposed in the following steps:
\begin{enumerate}
    \item Begin with the set of orders $(0,0,\ldots,0)$ for the $n$ variables (zero-order model) and compute BIC (equal to the variance of input noise).
    \label{BTS:step1}
    \item Increase the order by one for each variable separately. For example, for the first iteration starting with the set of orders $(0,0,\ldots,0)$ the $n$ candidate sets of orders are
    \[
    (1,0,\ldots,0), (0,1,\ldots,0), \ldots, (0,0,\ldots,1)
    \]
    \label{BTS:step2}
    \item Compute BIC for the $n$ dynamic regression models, one for each set of orders.
    \label{BTS:step3}
    \item Find the set of orders for which the BIC value is smaller than the BIC value of the previous step. If there are more than one such set of orders select the one with the smallest BIC value and go to next step. If no set of order has smaller BIC than the BIC from the previous step increase the lag by one and go to next step.
    \label{BTS:step4}
    \item If the current set of order is $(K_{\mathrm{max}},K_{\mathrm{max}},\ldots,K_{\mathrm{max}})$ then terminate, otherwise go to step~\ref{BTS:step2}.
    \label{BTS:step5}
\end{enumerate}
Upon termination, BTS delivers the current set of orders. Note that in step~\ref{BTS:step2} the order is increased only to the variables for which the current order is less than $K_{\mathrm{max}}$.

\section{Model order and parameter estimation methods \label{modmeth}}

We discuss here other methods for model order estimation as well as methods for the estimation of the parameters of the model.

\subsection{Order Estimation\label{ordest}}

\begin{figure*}[t!]
\centerline{\hbox{\includegraphics[height=8cm]{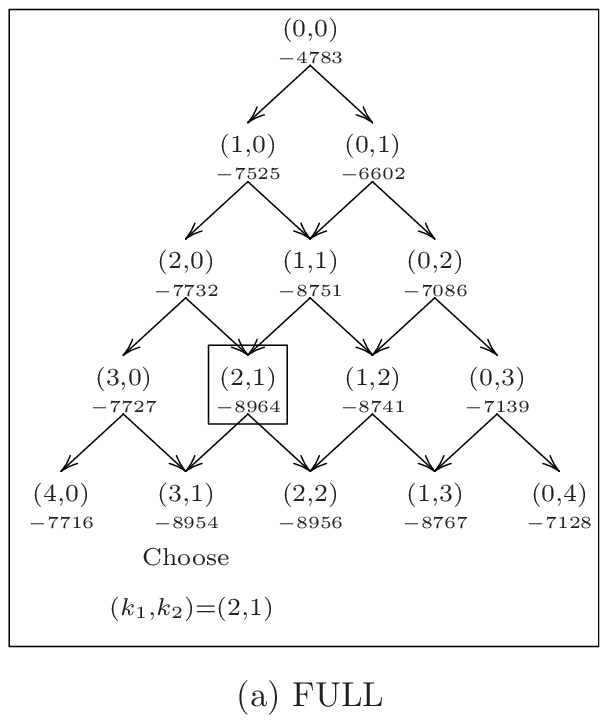}
\includegraphics[height=8cm]{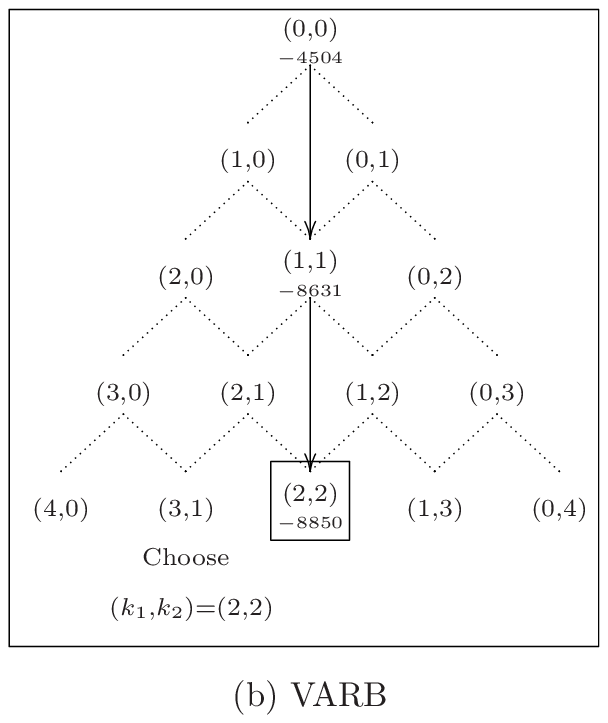}}}
\centerline{\hbox{\includegraphics[height=8cm]{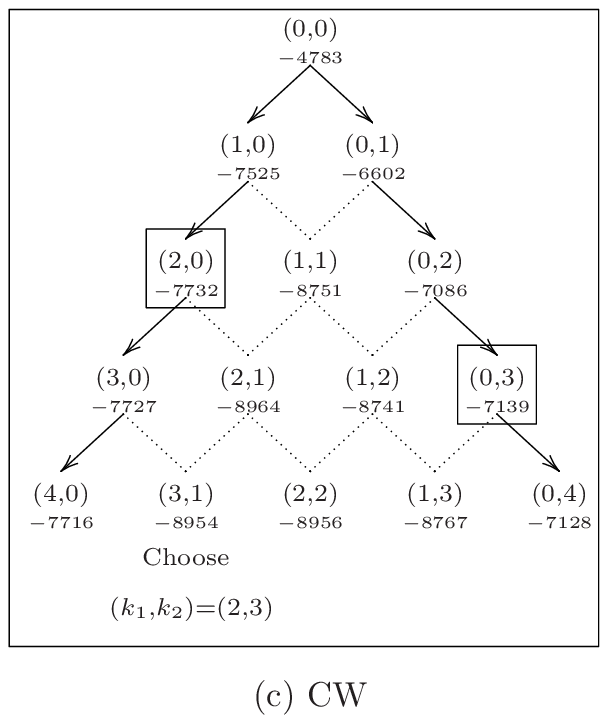}
\includegraphics[height=8cm]{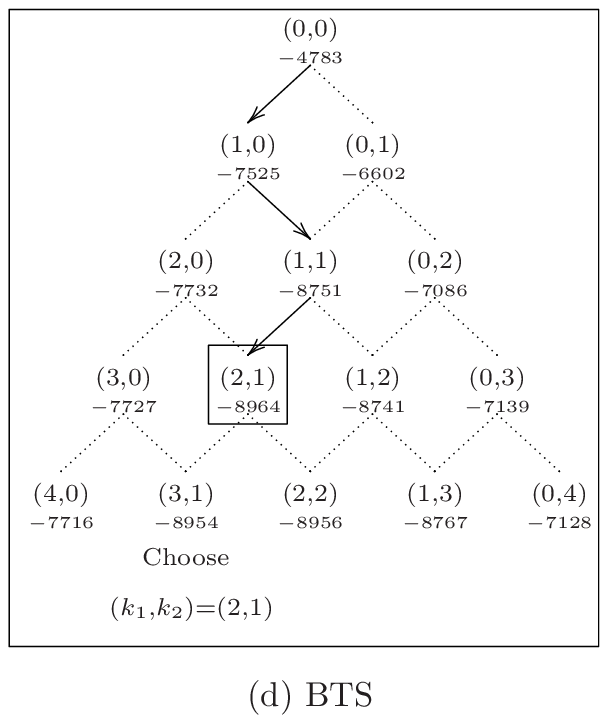}}}
\caption{Example of the order selection methods FULL, VARB, CW and BTS in (a), (b), (c) and (d), respectively. At each node the order vectors along with their corresponding BIC values are given. Solid arrows point to the order
vectors tested by each method for the determination of the optimum model in terms of
minimum BIC, which is marked in a square box.} \label{fig:methods}
\end{figure*}

In this study, we compare BTS to four other methods for estimating the orders of the DR model, listed below.
\begin{enumerate}
    \item Use of the same maximum order, i.e. $(K_{\mathrm{max}},K_{\mathrm{max}},\ldots,K_{\mathrm{max}})$ (MAX).
    \item Inspection of all possible combinations of model orders from
    $(0,0,\ldots,0)$ up to $(K_{\mathrm{max}},K_{\mathrm{max}},\ldots,K_{\mathrm{max}})$ (FULL)
    \item Optimum order of VAR($k$) for $k=1,\ldots,K_{\mathrm{max}}$ (VARB).
    \item Optimum orders selected from component-wise fits of $y_{1,t+1}$
    separately on each time series $y_{i,t}$ and lags up to $K_{\mathrm{max}}$ for $i=1,2,\ldots,n$
    (CW).
\end{enumerate}
In Fig.~\ref{fig:methods}, the steps that methods FULL, VARB, CW and BTS follow to obtain the optimum orders are illustrated for an example of a bivariate time series of the system DR(2,1) for $y_{1,t}$ and DR(0,1) for $y_{2,t}$ (not shown)
\begin{eqnarray}
y_{1,t+1}&=&0.7y_{1,t}-0.2y_{1,t-1}+0.5y_{2,t}+e_{1,t+1}, \nonumber\\
y_{2,t+1}&=&0.6y_{2,t}+e_{2,t+1},
 \nonumber
\end{eqnarray}
where $e_{1,t},e_{2,t}$ are normal iid.

The FULL method although best, since it inspects all possible models, is very time and
resource consuming when the number of time series is large. For example, for 16 time series and 3 lags
(0, 1 and 2) there are $3^{16}$ different models, that is a little more than 43 million
cases. It is obvious that this makes its application prohibitive for large systems. VARB regards the order of a VAR model and thus, if the component time series have different lag dependencies, VARB may
overestimate or underestimate the orders according to the strength of the
cross-correlations between the time series.

The CW method, commonly used in applications of DR models, determines the optimum order
(number of delays) for each variable independently \citep{Pena07}. Thus the same or similar dependence
forms of the response $y_{i,t+1}$ to different (delayed) variables may occur in the
selected model. To avoid this we propose in BTS an iterative method where at each step a delayed
variable that explains best the response is added to the current model.
CW often overestimates the orders because it does not take into account possible
cross-correlations. It may also underestimate the orders when the data sets are small or the
correlations between the variables are weak. A simple modification of CW would be to
select one variable that best fits the dependent variable, regress this dependent
variable to the chosen one and its lag, estimate the residuals and select the next
variable that regresses best these residuals. While this may work on large data sets, for
small time series with multi-collinearity this approach performs very badly, as we found from a pilot simulation study. The first selected
variable may explain information of the dependent variable that originates from other variables that relate to it and so the residuals obtained retain very small amount of information from the other variables, which
eventually terminates prematurely the search for further regressors on the residuals.

BTS attempts to render for the possible cross-correlations. The progressive scheme in BTS
is much faster than the exhaustive search scheme of FULL and it is intuitively sound
because variables are added to the model with respect to both their correlation to the
response variable and to their time proximity, something that we would expect to hold in
the context of time series.

\subsection{Parameter Estimation\label{parest}}

Depending on the selected orders $(k_{1},k_{2},\ldots,k_{n})$ of the DR model, the use of regularization in the estimation of model parameters may alter the model structure and reduce the degrees of freedom, i.e. the actual model coefficients to be estimated. Therefore we consider parameter estimation in addition to model order selection in order to assess the prediction efficiency of BTS and the other four methods for model order selection.

We assume that a model DR($k_{1},k_{2},\ldots,k_{n}$) is selected to be fitted to a time
series $\{y_{1,t}\}_{t=1}^{N}$. We consider a vector $\mathbf{b}$ of size
$K=\sum_{i=1}^{n}k_i$ containing all the coefficients of the lag polynomials in the DR
model. Let $\mathbf{X}$ be the lag matrix with rows
$$\tilde{\mathbf{x}}_t=[y_{1,t},y_{1,t-1},\ldots,y_{1,t-k_{1}+1},y_{2,t},\ldots,y_{n,t-k_{n}+1}]$$
for $t=\mathrm{max}(k_i),\ldots,N-1$. Assuming the one-step ahead fit for the response
vector of the variable $y_1$ is
$\mathbf{y}=[y_{1,\mathrm{max}(k_i)+1},y_{1,\mathrm{max}(k_i)+2},\ldots,y_{1,N}]'$, the
parameter vector $\mathbf{b}$ is estimated by the minimization of the error term $\mathbf
e$ in
\begin{equation}
 \quad\quad\quad\quad\quad\quad \mathbf{y}=\mathbf{X}\mathbf{b}+\mathbf{e}.
\label{eq1}
\end{equation}

The ordinary least squares (OLS) solution for $\mathbf{b}$ and regularizations of OLS can be expressed in terms of singular values decomposition (SVD) as follows (for a more thorough review and other regularization techniques see \citet{Kugiumtzis98,Jolliffe02}).
Suppose the SVD of the matrix $\mathbf{X}$, $\mathbf{X=U \Sigma V^\top}$ with
$\mathbf{U^{\top}U=I}_{N' \times N'}$, $(N'=N-\mathrm{max}(k_i))$,
$\mathbf{V^{\top}V=I}_{K \times K}$ and
$\mathbf{\Sigma}=\mathrm{diag}(\sigma_1,\sigma_2,\ldots,\sigma_K)$ a diagonal matrix with
components the singular values of $\mathbf{X}$ in descending order.
\begin{enumerate}
    \item The OLS estimate is
    $$\mathbf{b}_{\mathrm{OLS}}=\mathbf{V\Sigma^{-1} U^{\top}y}.$$
    This is the standard solution used for model fitting and it was used also in Section \ref{ordest}
    for the determination of the orders of the DR model.

    \item The Principal Components Regression (PCR) estimate restricts the OLS solution using only the first $q$ of the $K$ singular components \citep{Jolliffe02}
    $$\mathbf{b}_{\mathrm{PCR}}=\mathbf{V\Sigma^{-1} \Lambda_{\mathrm{PCR}}
    U^{\top}y},$$
    where $\mathbf{\Lambda_{\mathrm{PCR}}}$ is a diagonal $K \times K$ matrix with 1 at the first $q$ diagonal
    entries and 0 at the rest.

    \item The Partial Least Squares (PLS) estimate makes also use of a subspace of dimension $q\le K$ and is given in terms of SVD as \citep{Helland88}
    $$\mathbf{b}_{\mathrm{PLS}}=\mathbf{V\Sigma^{-1} \Lambda_{\mathrm{PLS}}
    U^{\top}y},$$
    where $\mathbf{\Lambda_{\mathrm{PLS}}}$ is a diagonal $K \times K$ matrix with
    components
    $$\lambda_i=1-\prod_{j=1}^{q}\big(1-\sigma_{i}^{2}/\theta_j\big),\ i=1,2,\ldots,K,$$
    where $\theta_j$ are the eigenvalues of
    $\mathbf{W}_{q}^{\top}\mathbf{X^{\top}XW}_{q}$ and the $\mathbf{W}_{q}$ matrix has as
    columns an orthonormal base of the space spanned by the vectors
    $$\{\mathbf{X^\top y,\big(X^\top X\big)X^\top y,\ldots,\big(X^\top X\big)}^{q-1}\mathbf{X^\top
    y}\}.$$

    \item The Ridge Regression (RR) (also called Tikhonov regularization) estimate \citep{Hoerl70}  involves a
    so-called ridge parameter $a$ and is defined in terms of SVD as
    $$\mathbf{b}_{\mathrm{RR}}=\mathbf{V\Sigma^{-1} \Lambda_{\mathrm{RR}}
    U^{\top}y},$$
    where $\mathbf{\Lambda_{\mathrm{RR}}}$ is again a diagonal matrix with
    components $$\lambda_i=\sigma_{i}^{2}/\big(\sigma_{i}^{2}+a\big).$$

\end{enumerate}

The methods PCR, PLS and RR aim at shrinking the space created by the rows of
$\mathbf{X}$ \citep{Lingjaerde00}. Their use is common in multiple regression when
multi-collinearity is present in the data, with PCR omitting components that have small
variability, PLS shrinking each component with regard to their predictive contribution
and RR shrinking them all by a percentage of their singular value.

We give a simple example for the problem arising in the case of multi-collinear time
series that illustrates the need for regularization. Let $\{y_{1,t}\}_{t=1}^{N}$ a time
series generated by an AR(1) model $y_{1,t+1}=\phi y_{1,t}+e_{t+1}$, where $e_t$ is iid,
and a second identical series, $y_{2,t}=y_{1,t}$. The two series are perfectly collinear,
and $y_{1,t}$ can be predicted equally good by, say, $\hat y_{1,t+1}={\phi \over 2}
y_{1,t}+{\phi \over 2} y_{2,t}$ or even $\hat y_{1,t+1}=100\phi y_{1,t}-99\phi y_{2,t}$.
Both these expressions are DR(1,1) models, and in fact both VARB and CW methods select
these orders. Since the columns of $\mathbf{X}$ are identical, $\mathbf{\Sigma}$ is
singular and $\mathbf{\Sigma}^{-1}$ cannot be computed. Even if the two time series were
not perfectly collinear, but just strongly collinear, $\mathbf{\Sigma}$ would be near
singular (very badly conditioned) and the OLS estimation would be numerically unstable.
By multiplying $\mathbf{\Sigma}^{-1}$ with the appropriate $\mathbf{\Lambda}_\bullet$
matrix ($\bullet$ denotes PCR, PLS or RR) the effect of collinearity can be corrected.

\subsection{Regularization parameter selection\label{regparsel}}

For the regularization methods we need to determine appropriate values for their
parameters, i.e. $q$ for PCR or PLS and $a$ for RR. For the optimal selection of these
parameters we use a 10-fold cross-validation criterion \citep{Breiman92,Kohavi95}. The
sample of length $N-\mathrm{max}(k_i)$ (the rows of $\mathbf{X}$) is split into 10
consecutive segments of equal length. The parameter vector $\mathbf{b}_\bullet$ for a
given model, regularization method and regularization parameter is estimated from a
subsample consisting of 9 of the segments and the fitted model is applied to the excluded
segment to predict the values $\hat y_{i,t+1}$ in it. We do the same for all segments
until we have predictions for all the values of the data. The measure of goodness of fit
is the sum of squared errors of the fit
$$\mathrm{SSE}=\sum\limits_{t=\mathrm{max}(k_i)}^{N-1} \big(\hat y_{i,t+1}- y_{i,t+1}
\big)^2.$$ We compute SSE for a range of values of the regularization parameter and the
optimal is the one giving minimal SSE.

For PCR and PLS, the regularization parameter $q$ takes only limited integer values and
the inspection is performed for all values of $q=1,2,\ldots,K$. For RR the process is a
bit more complicated because $a\in \mathbb{R}^{+}$ and we employ the following search
scheme. We compute the $\mathrm{SSE}$ for 11 values of $a$ in the interval
$[0,\sigma_1]$, i.e., $a\in\{0, 0.1\sigma_1,0.2\sigma_1\,\ldots,1\sigma_1\}$ and locate
the one giving the minimum $\mathrm{SSE}$, let it be $a'$. We set the new interval
starting at the value preceding $a'$ in the set of 11 values and ending at the value
following $a'$ in the set (if $a'$ is the first or last value in the set, then $a'$ is
the respective edge of the new interval). We repeat the process until the relative change
of $\mathrm{SSE}$ in two consecutive steps is very small (less than $10^{-6}$ in our
applications), where the relative change from ${\mathrm{SSE}(a_{\mathrm{previous}})}$ to
${\mathrm{SSE}(a_{\mathrm{current}})}$ is
$${{\mathrm{SSE}_{\mathrm{RR}}(a_{\mathrm{previous}})-\mathrm{SSE}_{\mathrm{RR}}(a_{\mathrm{current}})} \over {\mathrm{SSE}_{\mathrm{RR}}(a_{\mathrm{previous}})}}.$$

\section{Monte Carlo Simulations\label{mocasi}}

We evaluate the proposed method BTS for order selection in comparison to the other four methods. In the simulation study, we evaluate also whether additional constraint to the coefficients of the lagged variables, as set by the three regularization techniques, can be of any benefit towards better prediction. The regularizations are expected to have a large effect when the model has a fixed large order, as for the MAX order selection method, and much less effect when the orders are small and close to the real orders, as expected for the other order selection methods.

\subsection{Monte Carlo Setup\label{mocase}}

We study linear systems with feedback and multi-collinearity of varying
strength. We consider small time series of length $N=100,200,400$, since these are mostly
affected by the aforementioned data conditions. We split each time series to a training
set of the first $3N/4$ samples and a test set of the rest samples.
On the basis of the training set we determine the optimum model for each of the 20 combinations of 5 methods of order estimation and 4 methods of parameter estimation.
We apply each selected model on the data of the test set and calculate the Normalized Mean Squared Error (NMSE) of one-step ahead predictions
\begin{equation}\nonumber
\mbox{NMSE}={{\sum\limits_{t} \big(y_{i,t+1}-\hat{y}_{i,t+1}\big)^{2} \over
\sum\limits_{t} \big(y_{i,t+1}-\bar{y}_{i}\big)^{2}}}={\sigma_{\hat{y}_i}^2+b^2 \over
\sigma_{y_i}^2}={\sigma_{\hat{e}_i}^2 \over \sigma_{y_i}^2}, \label{eq:NMSE}
\end{equation}
where $\hat{e}_{i,t+1}=y_{i,t+1}-\hat{y}_{i,t+1}$ are the prediction errors,
$\bar{y}_{i}$ is the mean of the actual values of $y_{i,t+1}$ over all target times $t$,
$b^2$ is the bias introduced by the regularization technique ($b^2=0$ for OLS) and
$\sigma^2_\bullet$ is the variance of $\bullet$.

For each system we perform 1000 Monte Carlo realizations with the help of the Matlab
computation environment and compute the average NMSE, for all 20 methods of order and model
estimation. Comparing different methods for optimum model identification is not an easy task. We
can't say that a method is best just because it detects best the real model that
generated the time series in a Monte Carlo study (finding the correct model order most
times out of the 1000 realizations, or giving most accurate estimates of parameters),
since there may be more than one equivalent models \citep{Castle11}. We deem that the average NMSE is a
good indicator of the performance of each method. Equivalent model representations will
have equivalent prediction errors and thus NMSEs, so the difference in the average NMSEs
of two methods will depend solely on the non-equivalent realizations results. The failure
of a method in providing an appropriate model will be indicated by a significant increase
of the average NMSE.

Furthermore, we need to asses the significance of the difference in the prediction
performance. For this we use the Diebold-Mariano test \citep{Diebold95}. The null
hypothesis for this test is that the two models under investigation have the same
out-of-sample prediction accuracy. For each of the 1000 realizations we apply the test at
a 5\% significance level on all pairs of methods and record the number of rejections. For
a pair of equivalent methods we expect to have 50 rejections by chance and from these
about 25 times the first method to give better predictions and 25 the second. Thus we
consider two methods to be evidently different if in addition there are at least 50 more
rejections, a number chosen arbitrarily to signify some power of the test. In the summary results, when presenting the
average NMSEs of all methods we mark the one with the minimum NMSE and those that are
deemed as not different to it according to the Diebold-Mariano test.

For the evaluation of each $j$ method of model order selection and parameter estimation,
$j=1,2,\dots,20$, on a number of different settings (different systems and variables) we
use an efficiency indicator (score) $\mathrm{S}_j$. For each case $i$, $i=1,2,\ldots ,M$
we compute NMSE on the test set and the noise to signal ratio
$\sigma_{e_{i}}/\sigma_{y_{i}}$, which corresponds to the square root of the reference
NMSE (e.g., the selected model is the real one). We define the score as
\begin{eqnarray}
 \nonumber
 \hspace{45pt} \mathrm{S}_j &=& \sum\limits_{i=1}^{\mathrm
M}{\big(\big(\sigma_{e_{i}}/\sigma_{y_{i}}\big)^2-
\big(\sigma_{\hat{e}_{i}^{j}}/\sigma_{y_{i}}\big)^2\big)^2 \over
\big(\sigma_{e_{i}}/\sigma_{y_{i}}\big)^2} \\
   &=& \sum\limits_{i=1}^{\mathrm
M}{\big(\sigma_{e_{i}}^2-\sigma_{\hat{e}_{i}^j}^2\big)^2 \over
\big(\sigma_{y_{i}}\sigma_{e_{i}}\big)^2}. \label{sscore}
\end{eqnarray}
The superscript $j$ of $\hat{e}_{i}^{j}$ denotes that these errors where obtained by
using method $j$. The score $\mathrm{S}_j$ measures the degree of approximation of the
real model (on a number of cases) with model $j$ by normalizing the prediction error with
respect to the variance of the time series and the noise level.

\subsection{Results\label{mocares}}

\subsubsection{A 4-d VAR model of order 2 \label{mocvar2}}

The first system is a VAR of order 2 in 4 variables, given by
\begin{eqnarray}\label{var2mod}
\nonumber
  \mathbf{y}_{t+1} &=&  \begin{small} \left( \begin{array}{cccc}
0.3 & 0  & 0   &  0 \\
0.4 & 0  & 0.7 & -0.9 \\
0.7 & -0.6 & -0.5 & 0\\
0.3 & -0.2 & 0 & -0.4 \end{array} \right) \end{small} \mathbf{y}_{t}+ \\
    &+& \begin{small}\left(\begin{array}{cccc}
-0.5 & 0  & 0   &  0.2 \\
0 & -0.3  & -0.1 & 0 \\
0 & -0.1 & 0.2 & 0.4\\
0 & 0 & 0 & 0.6 \end{array} \right)\end{small} \mathbf{y}_{t-1}+\mathbf{e}_{t}
\end{eqnarray}
with $\mathbf{e}_{t}\sim \mathrm{N}(\mathbf{0},0.1\cdot\mathbf{I})$ ($\mathbf{0}$ a
vector of 4 zeros and $\mathbf{I}$ the $4\times 4$ identity matrix). Given the zero
entries in the coefficient matrices the VAR system can be decomposed to 4 DR systems one
for each variable, namely DR(2,0,0,2), DR(1,2,2,1), DR(1,2,2,2) and DR(1,1,0,2)
respectively. Since some of the non-diagonal entries of the coefficient matrices are
zero, feedback is present between the time series for each DR. For example, in the first
DR system the dependence of $y_{1,t+1}$ on $y_{4,t-1}$ can in turn be explained by the
dependence of $y_{4,t-1}$ on other variables, including $y_{1,t-2}$. Thus through
$y_{4,t-1}$ there is a (feedback) dependence of $y_{1,t+1}$ to $y_{1,t-2}$ in addition to
the dependence on $y_{1,t}$ and $y_{1,t-1}$ being present in DR. The most frequently
selected orders by the four methods (i.e. excluding MAX) for $K_\mathrm{max}=5$ along
with their frequencies are given in Table \ref{tab:var2freqs}.

For small samples
($N$=100) all methods that can give unequal orders do not single out a particular model
order and the results vary from one time series to another. As the sample size increases,
so does the frequency of the most selected model orders for FULL and BTS while CW shows
quite large spread of the selected orders. It is notable that CW never manages to pick
the real orders with highest frequency, whereas BTS (alike FULL) does this in most of the
cases. VARB rather constantly estimates order (2,2,2,2) even for small samples (only in
about 30 of the 1000 realizations a different order is selected).

The average NMSEs for the prediction with the 20 joint methods of order selection and
parameter estimation are given in Table \ref{tab:var2nmse}.

For each variable and sample
size the best methods found from the Diebold-Mariano test to have equal predictive
accuracy are marked with an asterisk. The RR regularization improves slightly the NMSE of
FULL, VARB and BTS for small $N$ and somewhat less for larger $N$, whereas PLS and PCR do
not have any significant effect. These three methods converge for large samples with VARB
being overall best indicating that the best fit guaranteed by FULL does not always yield
best predictions. CW and MAX perform poorly especially for small sizes, and for time
series 1 and 2, with CW being the worst of the two. The prediction performance of the
modeling methods is not consistent across the 4 time series and depends highly on the
structure of the particular DR system.

\subsubsection{A 4-d VAR model of order 2 with correlated errors\label{mocvar2corer}}

This system is identical to the one given by eq.~\ref{var2mod} with the components of
$\mathbf{e}_{t}$ being correlated according to the correlation matrix
$$\mathbf{R}= \begin{small} \left( \begin{array}{cccc}
1 & 0.6  & -0.1   &  0.2 \\
0.6 & 1  & -0.3 & 0.4 \\
-0.1 & -0.3 & 1 & 0\\
0.2 & 0.4 & 0 & 1 \end{array} \right) \end{small}.
$$
The most frequently selected orders by the four methods (i.e. excluding MAX) for
$K_\mathrm{max}=5$ along with their frequencies are given in Table
\ref{tab:var2corerfreqs}, and the average NMSEs for the prediction with the different
methods of order selection and parameter estimation are given in Table
\ref{tab:var2corernmse}. The results are overall similar with the case of uncorrelated
input noise. The NMSEs are slightly higher for all methods, a fact that indicates that
the correlation in the input noise vectors hinders the prediction process, but does not
affect the methods themselves.




\subsubsection{Multiple DR systems \label{moc81DR1133}}

We assess now the prediction performance of the methods on multiple DR systems by means
of the efficiency scores. We consider bivariate time series, where the first is generated
by a DR($\mathrm{k}_{11}$,$\mathrm{k}_{12}$) system, and the second by a
DR($\mathrm{k}_{21}$,$\mathrm{k}_{22}$) system with $\mathrm{k}_{ij} \in \{1,2,3\}$ and
${e}_{t}\sim \mathrm{N}(0,0.1)$. We have $3^4=81$ different systems and since they are
bivariate we have 162 DR systems. The parameters for these systems were picked randomly
but constrained to give stationary time series and again 1000 realizations were generated
for each system. Results on individual systems are not very useful since they vary much.
The efficiency scores defined in eq.~\ref{sscore} were computed for M=162 DR systems and
are given in Table \ref{tab:Sj}.

For $N$=100 VARB performs best, even better than FULL that due to the small sample size
and the use of the BIC criterion omits parameters that turn out to contribute in
prediction. As sample size increases, FULL becomes best and for $N$=400 BTS becomes
second best. Again PLS and PCR do not improve FULL, VARB and BTS, while RR gives a
marginal improvement of VARB. CW is the worst by far for all sample sizes and
regularization techniques do not improve it either. The MAX method performs also quite
badly with RR regularization giving some improvement. For large $N$ all methods except CW
seem to converge.


\subsubsection{A system of 8 multi-collinear time series \label{moc8ts}}

Our previous results were on multivariate time series with inter-component feedback. Now
we focus on time series with cross-correlation by assigning direct dependencies among
them. A system of 8 multi-collinear time series is created in the following way. First 7
AR(1) time series are created as $y_{i,t+1}=a_{i}y_{i,t}+e_{i,t+1}$ with $e_{i,t}\sim
\mathrm{N}(0,0.1)$ and $a_i$ takes the value -0.76, -0.89, 0.59, 0.62, 0.87, -0.72, -0.61
for $i=1,\ldots,7$, respectively. We create collinear time series by superimposing to
each of the last 6 time series the first time series (being a sort of ``common
component'') multiplied by a coefficient c
$$x'_{i,t}=y_{i,t}+\mathrm{c}y_{1,t},$$ for i=2,3,\ldots,7 and c=0,0.5,1,2. The coefficient c controls the
strength of the collinearity between the time series. Finally we create the 8-th time
series from the mean of time series 2, 3 and 4
$$x'_{8,t}={1 \over 3}\big(x'_{2,t}+x'_{3,t}+x'_{4,t}\big).$$

We post the problem of predicting time series 8, $x'_{8,t}$, from all 8 time series. The
results for all methods using $K_{\mathrm{max}}=3$ are given in Table \ref{tab:C8}. For
the sake of clarity we omit methods that do not indicate substantial differences, like
the use of regularization techniques for method FULL and when regularizations give
similar results we present only the best one. We note though, that RR gives almost always
at least marginally best results when there is strong collinearity (c=1 or 2), as we also
observed in the case of the 4 dimensional VAR(2) system.

Generally MAX method gives bad results since the number of parameters to be estimated are
many, relative to the sample size, particulary when $N$=100 where for c=0 NMSE is 1, meaning that the model involving
24 parameters gives as good predictions as the mean of the data. The use of RR
regularization in the MAX method improves significantly the predictive ability (PLS and
PCR have the same effect but at a smaller extent) and in the presence of strong
collinearity (c=2) and for small sample size its performance is even comparable to FULL, being
generally the best. However its improvement with the sample size is small compared to the
other methods. Method CW without regularization performs quite good for large sample
sizes and strong collinearity. For weak or no collinearity FULL performs best for small
samples with BTS following and VARB performing a bit worse. As sample size increases
methods VARB and BTS converge to the performance of FULL. As c increases the convergence
of the three methods can be observed even for small $N$.


\subsubsection{A second system of 8 multi-collinear time series \label{moc8ts2}}

As an extension of the previous case we create a similar system using an AR(2) system
instead of AR(1) for the ``common component''. Thus time series $\{y_{1,t}\}$ is
generated by $y_{1,t+1}=-0.76y_{1,t}-0.60y_{1,t-1}+e_{1,t+1}$ and the correlations of the
time series extend further back in time. The results on prediction are given in Table
\ref{tab:C82}. Again we show only cases showing particular features of the methods.

Methods FULL, BTS and MAX behave similarly to the previous case with BTS again giving equal good predictions to FULL. CW performs quite worse than before and for strong collinearity needs regularization to have good results for small $N$. Method VARB cannot catch up with FULL and BTS in this case and we explain this as follows. Unlike the previous system, this system involves significant cross-correlations for larger delays, but the constraint of small data size prevents VARB from reaching the larger ``true'' optimum
model order. Thus VARB chooses almost always model order 1 as it did also for the first
system, where the correlations were actually of order 1 and therefore the model performed
well.


\section{Application to real data\label{eegdata}}

\subsection{EEG data}
Our data sets are three 25-channel recordings of scalp EEG from an epileptic patient with
generalized tonic clonic seizure with sampling time 0.01 sec (the data were provided by the Oslo University Hospital). We use two records of
duration 1 hour (360000 data points), the first from 4 up to 3 hours before seizure
(early preictal period), the second 1 hour prior to seizure (late preictal period) and a
third 15 minutes long record (90000 data points) during and after the seizure (ictal
period). Our goal is to check the predictive ability of the multivariate modeling
methods on the different periods and use it as a measure for discerning between them.
This is achieved with the help of Receiver Operating Characteristic (ROC) curves analysis
and the Area Under Curve (AUC) statistic \citep{Zweig93,Fawcett03}. Values of AUC near
0.5 indicate that there is no distinction, while values near 1 means that we have
complete separation.

Each one of the records is split into windows of 4 seconds duration (400 data points).
For each one of the 25 time series a dynamic regression model on all 25 time series is
estimated with each method on the first window. Then the one-step ahead predictions of
the model on the subsequent window gives the first NMSE. Then the second window is used
for model estimation and predictions are made on the third part and so forth. Thus for
each channel we have 899 NMSE values for the first and second record and 224 for the
third one. We omit the use of FULL method since the large number of time series prohibits
its application. For $K_{max}=3$ that we use, there are $4^{25}\approx 10^{15}$ different
models. We also omit the use of PCR estimation of the model parameters because its
performance is always worse than the similar to it PLS estimation. We compute the values
of AUC taking as samples the set of NMSE values over the three records and compare the
early preictal with the late preictal periods and the late preictal with the ictal.

\begin{figure*}
 \centerline{\includegraphics[width=17cm]{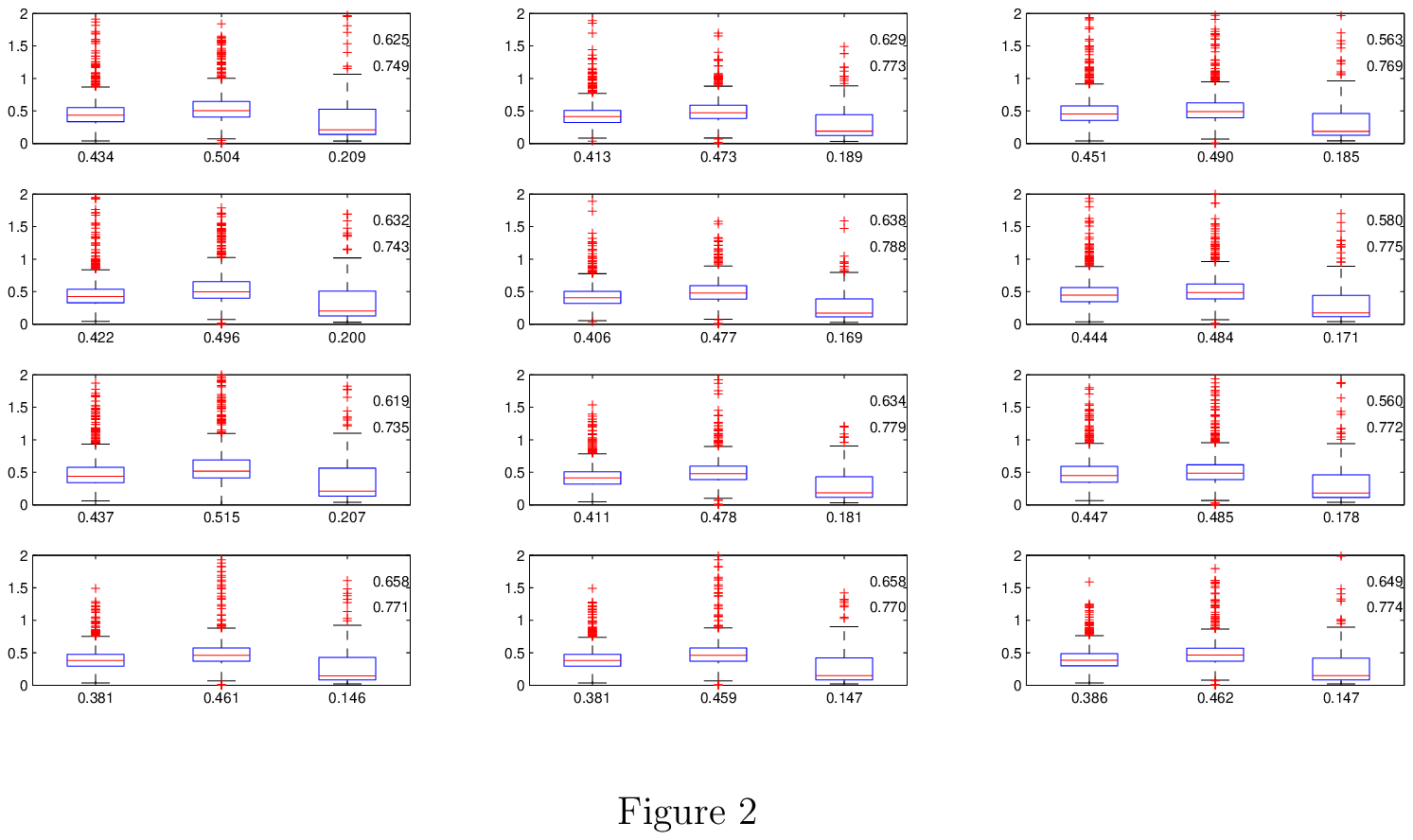}}
\caption{Each panel shows the boxplots of the NMSE values for channel 9 of the EEG data at the three records:
early preictal (left), late preictal (center) and ictal (right). Under each boxplot is the median NMSE. The panels are organized
as follows: the row sequence is VARB, CW, MAX and BTS; the column sequence is OLS, PLS
and RR. In the inset are the values of the AUC for comparing between record 1-2 (upper)
and 2-3 (lower).}
  \label{fig:channel9}
\end{figure*}

In Fig. \ref{fig:channel9} we give the results for one channel. Comparing
the methods for model selection with OLS parameter estimation (first column of boxplots)
we see that BTS has the smallest median NMSE for all three records, while the other
methods give, more or less, similar NMSEs. We opt for the use of median NMSE instead of
average NMSE because our data are heavily contaminated with artifacts and thus the distribution of NMSE is heavily right-skewed. With regard to the parameter estimation methods, for BTS there is no significant change of NMSE with either PLS or RR. For the other three methods, PLS decreases NMSEs for all records, whereas RR
increases NMSE for the first record and decreases it for the other two records. This is
possibly an indication that there is change in the dynamics of the underlying mechanism
producing the time series. The values of AUC are slightly higher for BTS and for the
other methods with PLS estimation of model parameters showing that there is a slight
increase of the discriminating efficiency.

Again the results of NMSE differ depending on the channel. To account for this, we choose
to compute the percentage difference on median NMSEs of the modeling methods with
reference to the BTS with OLS estimation of parameters and average it over all channels,
$${\mathrm{NMSE}_\bullet-\mathrm{NMSE}_{\mathrm{BTS_{ols}}} \over
\mathrm{NMSE}_{\mathrm{BTS_{ols}}}}*100$$ where $\bullet$ denotes any of the other
methods. The results are given in Table \ref{tab:AverMedianNmseInc}. There is no
significant change for BTS with regard to regularization. On the other hand, the NMSE of
the other methods ranges from 5.5\% (CW with PLS on record 2) to 51.3\% (VAR with OLS on
record 3) larger than that of BTS. The overall results are similar to the ones from the
single channel. Again PLS improves the performance of VAR, CW and MAX and RR performs
worse than OLS only on the first record and moderately on the other two. It is evident
that the overestimation of the model orders from VAR and CW methods (and MAX as well)
results to bad predictions and regularization corrects this only partially.


Regarding the discrimination of different epileptic phases with the NMSE as measure and
the difference between the methods, the averaged AUC values over all channels are given
in Table \ref{tab:AverAUC}. We see that although there is significant decrease of the
NMSEs for BTS, this does not contribute significantly toward a better discrimination.
VARB and MAX without regularization have smaller values for AUC on both record
comparisons than BTS and only with PLS regularization they manage to catch up with BTS.
RR worsens the results for the first comparison and slightly improves for the second, but
PLS performs better.


\subsection{MSCI data}

The second data set is the Morgan Stanley Capital International's (MSCI) market capitalization weighted index of 23 developed markets in North America, Europe, and the Asia/Pacific Region. It is calculated with the help of the equities values of companies representative of the market structure. The data set comprises of 1300 daily returns (first differences of the logarithms of the indices) for each market in the period 5/3/2004-5/3/2009, excluding weekends and holidays. The original set is split into 5, roughly, 1-year periods consisting of 260 data points. Each period is used for the estimation of the model with the different methods and then the NMSE of the model is calculated on the following one-year period. Thus for each market and method 4 values of NMSE are obtained. The maximum lag used is 3 indicated by the significant cross-correlations between the time series.

\begin{figure*}[t!]
\centerline{\hbox{\includegraphics[height=5cm]{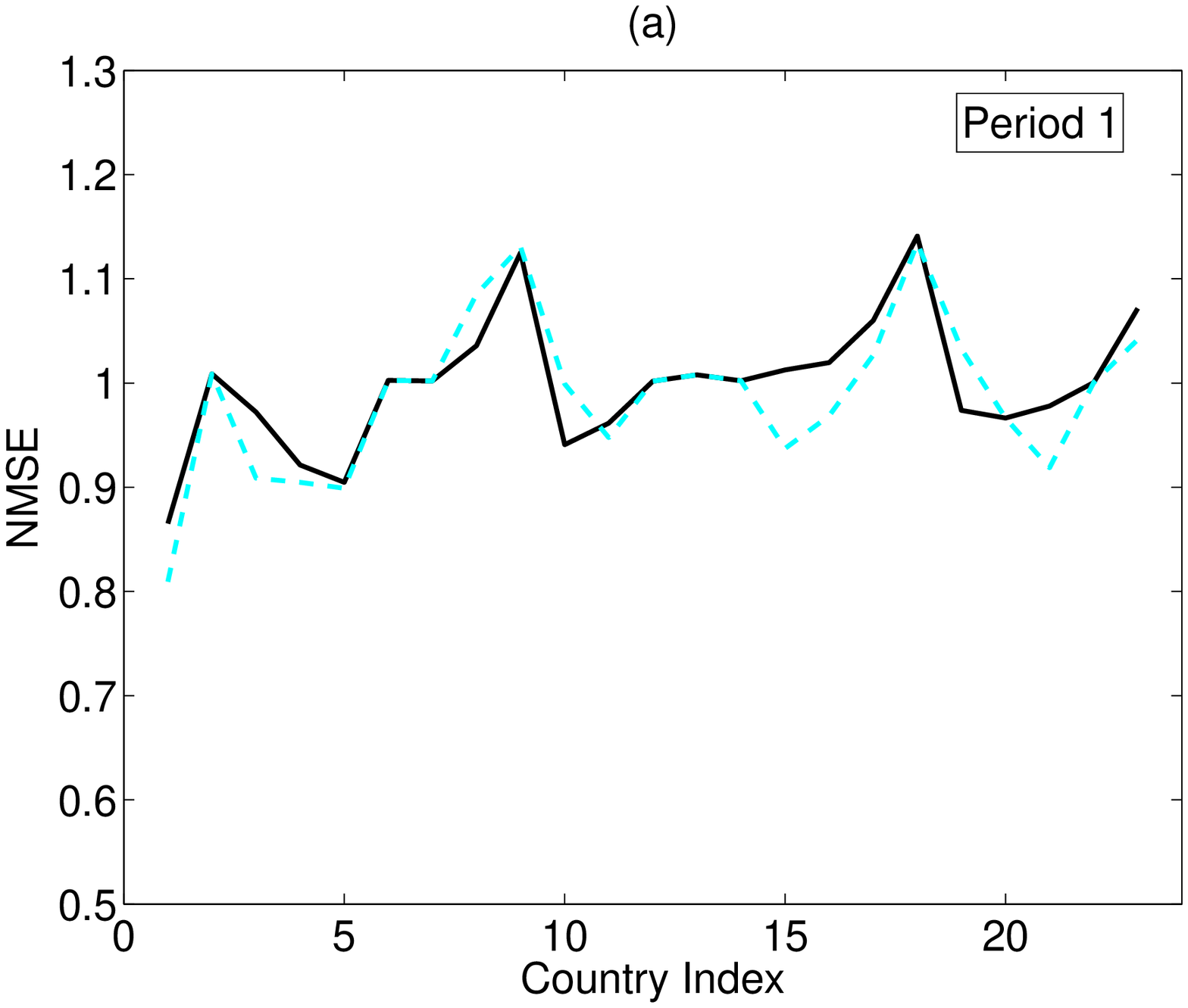}
\includegraphics[height=5cm]{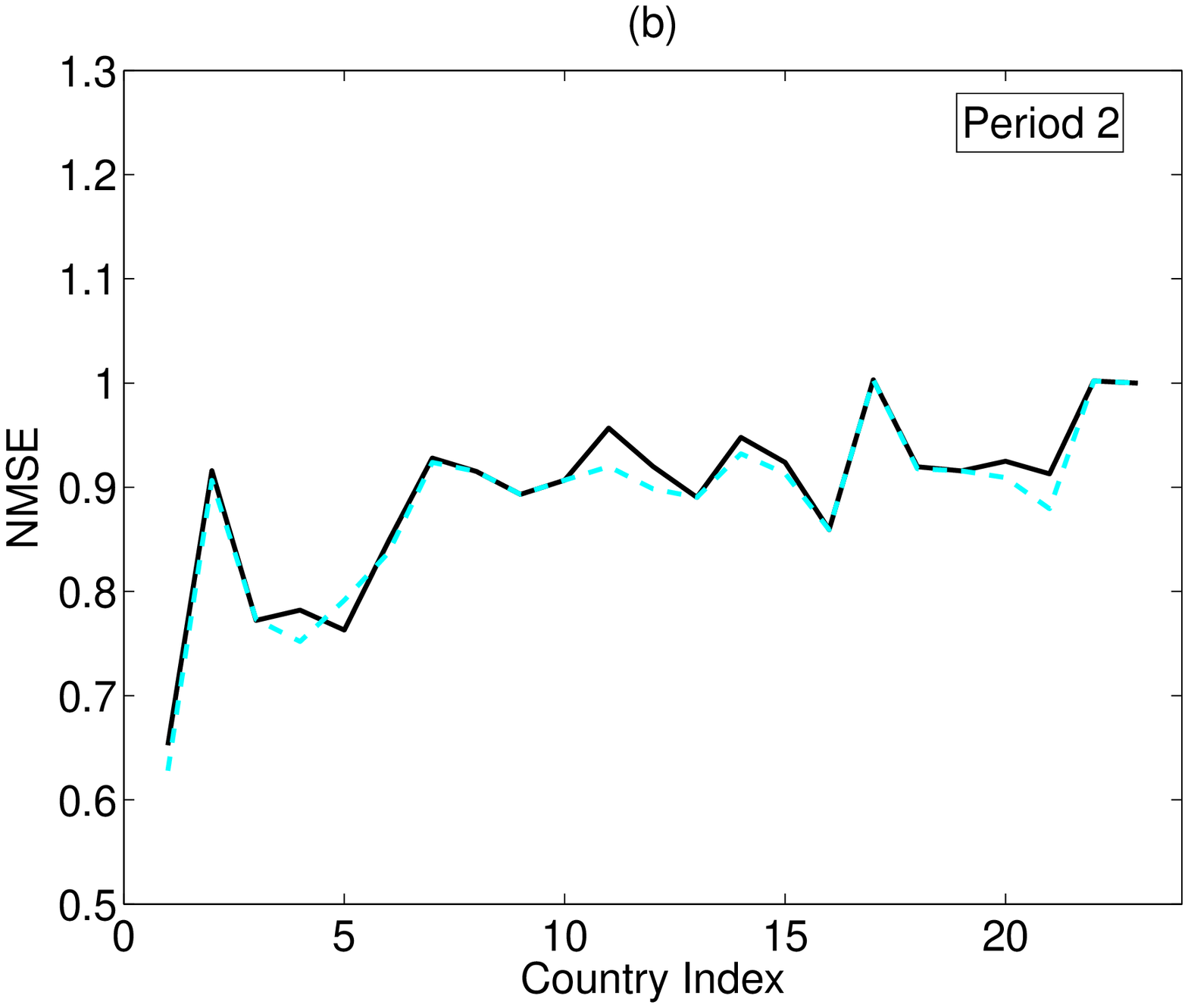}}}
\centerline{\hbox{\includegraphics[height=5cm]{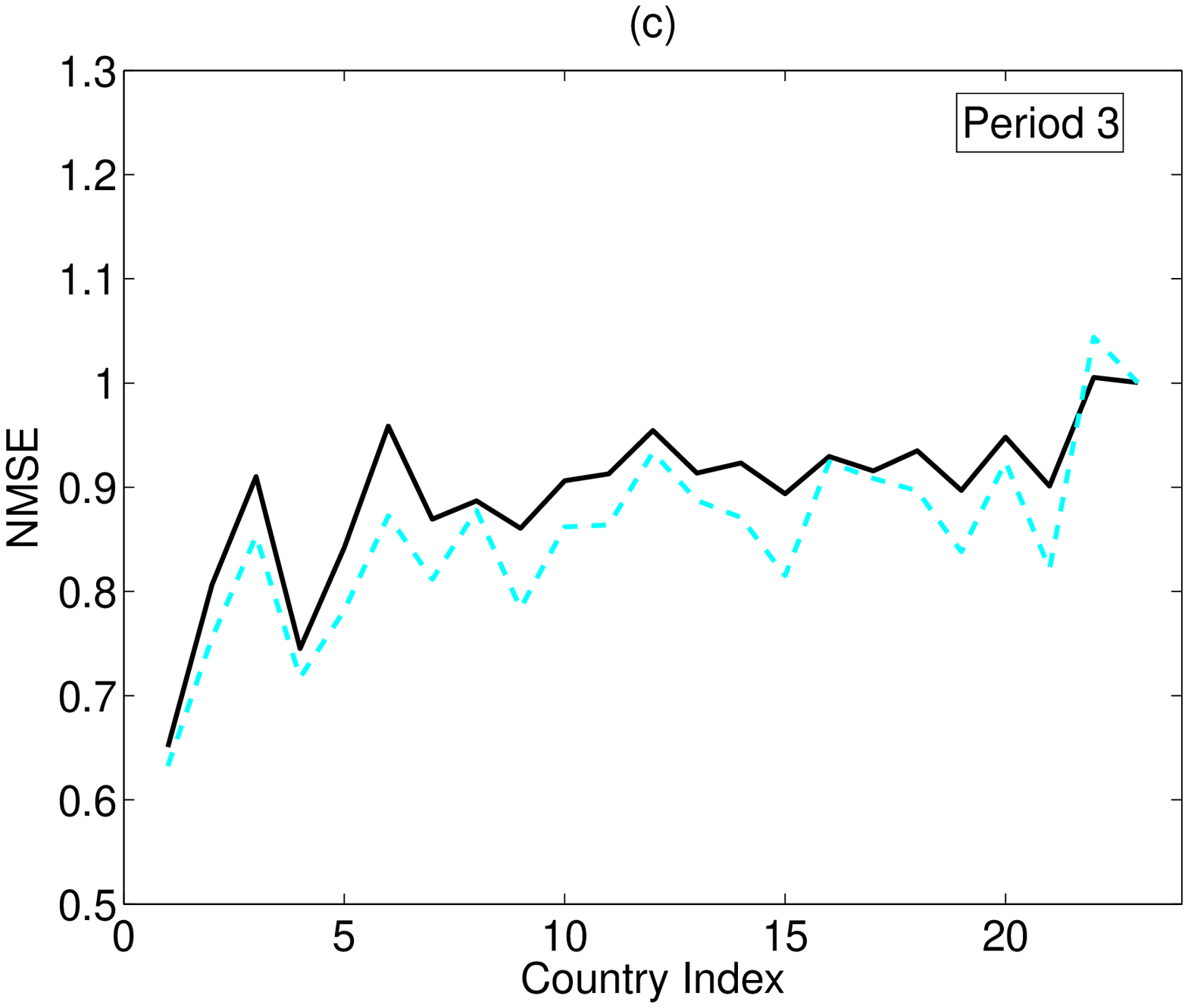}
\includegraphics[height=5cm]{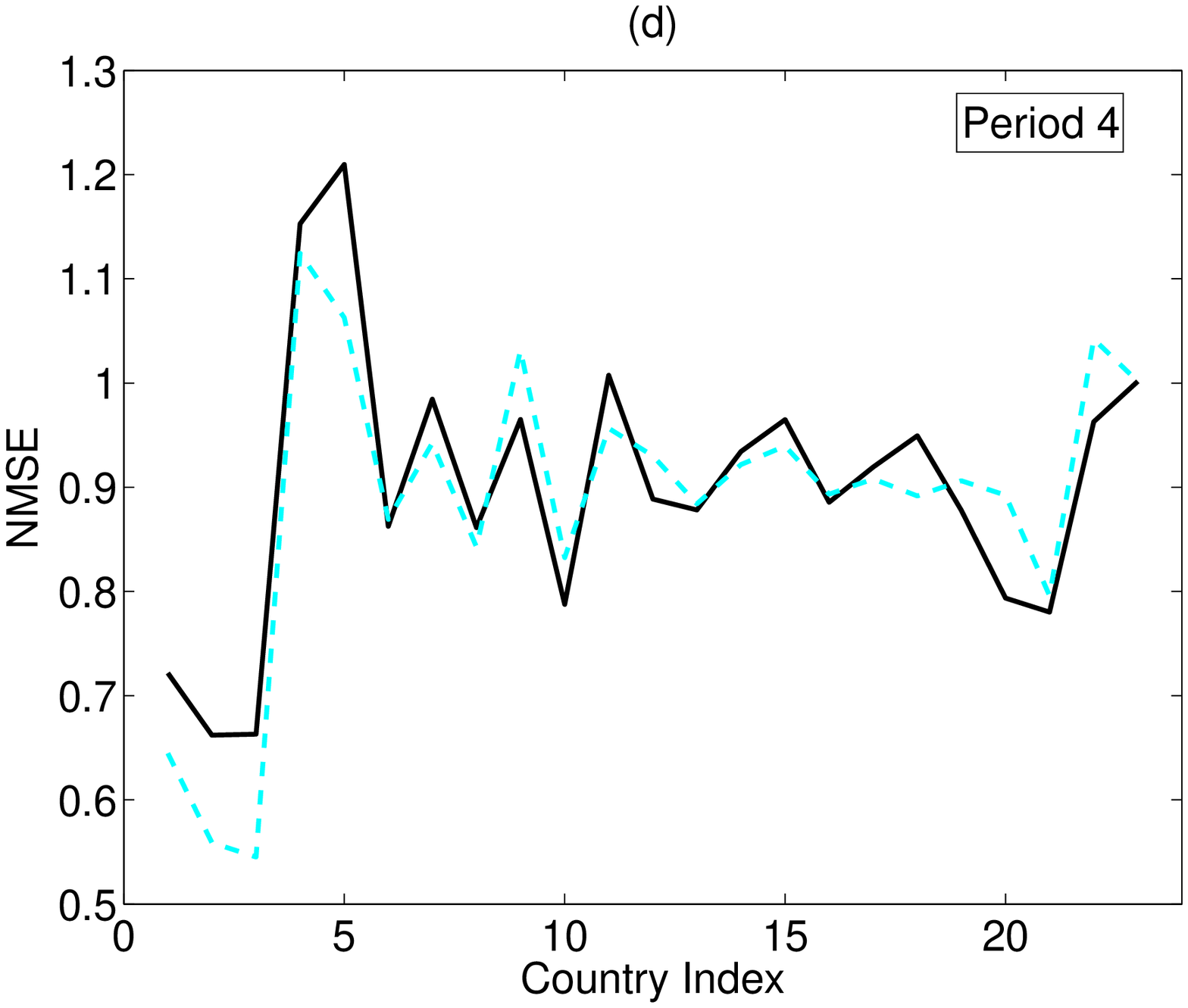}}}
\caption{NMSE for the 4 yearly periods and two methods of model estimation for the MSCI data. Each subfigure shows a different period and the two lines correspond to the two methods. Solid black line is CW and dashed gray (cyan online) is BTS, both with OLS parameter estimation. On the $x$-axis is the country index as shown in Table \ref{tab:MSCIDEV}.} \label{fig:MSCIper}
\end{figure*}

Figure \ref{fig:MSCIper} shows the NMSE values for all markets and period for the methods CW and BTS with OLS estimation of the model parameters. We see that the NMSE of BTS is lower or in par with that of CW. In the third period in fact BTS performs better for almost all countries. The average NMSE for each market over the 4 periods and for selected methods is shown in Table \ref{tab:MSCIDEV}, along with a total average over all markets. The best NMSE is given by BTS with RR parameter estimation (0.900) and is very close to BTS with OLS (0.904). CW with RR follows (0.909), while CW with OLS is fourth with somehow higher value (0.922). VAR and MAX with OLS fail completely (1.002 and 1.186 respectively) due to the large number of time series and small data length. Method VAR indicates zeroth order models for all cases, while MAX has stability problems on the parameter estimation and only with RR regularization manages to give results near the ones given by the other methods (0.929). All best performing methods (BTS with OLS, BTS with RR, and CW with RR) predict worst the two North American markets and best the Asian/Pacific markets, with UK giving the lowest NMSE (0.854 for BTS with OLS) among all European markets.


\section{Discussion\label{discus}}

The proposed BTS method for the selection of the order of a dynamic regression (DR) model forms the set of lagged variables progressively, starting from zero lag (order) and increasing one lag at a time. The lagged variables are selected according to their predictive relevance to the response, supplementary to this of the already selected subset of lagged variables. The time order in the selection of the lagged variables is intuitively sensible, because in real world data dependencies and correlations decrease as we go back in time. Also BTS is conservative since the model contains minimal redundant information, and computational inexpensive in its implementation.

We compared BTS with other, commonly used in practice, methods of order selection with and without regularization techniques in the model parameters estimation. The selected methods are quite basic and the performance of BTS compared to more sophisticated ones remains to be seen. Popular methods of subset selection and/or regularization used in multiple linear regression, such as forward stagewise regression \citep{Hastie08}, genetic algorithms \citep{Holland92}, least absolute shrinkage and selection operator (LASSO) regression \citep{Tibshirani96} or elastic net regularization \citep{Zou05} could be used in DR model selection. However, it should be noted that the application of these methods is not always straightforward due to the specific nature of time series data. An initial approach of the problem based on forward stagewise regression produced very bad results, probably due to the autocorrelation in the residuals.

BTS has in-built regularization, because only the relevant or effective lags are chosen, a kind of internal process of projecting the full space to the axes with best predictive relevance. Indeed the simulation study showed that regularization in the estimation of the parameters of the DR model determined by BTS had no effect on the predictive power of the model. The results on the other methods for order selection showed that the regularization improved prediction only when the order was overestimated. For small sample sizes, the order selection methods tend to underestimate the order of the model, so there is loss of information that cannot be compensated by any regularization. The regularization techniques have significant effect only in the case of a fixed large order (the scheme denoted as MAX), where there is no loss (but rather surplus) of information, but even so the effect is poor compared to other methods with no regularization, like BTS or VARB (the latter being the VAR order estimated by the BIC criterion).

The data conditions of feedback and multi-collinearity that we dealt with in the Monte Carlo simulations are rather typical in applications and the consensus is that when the data sets are large enough these cases do not pose a problem in modeling and prediction. However, this does not hold also for small sample sizes and we showed that the order selection methods give varying results with regard to the number of
observed time series and the data conditions. For different conditions of feedback and multi-collinearity, the predictions with BTS were consistently among the best and close to the predictions of the best order (as found by the exhaustive search). VARB performed well at some of the settings but failed when there was strong multi-collinearity, and the CW method (estimates using BIC optimal order for each variable separately) gave generally worse predictions than BTS.

For the real EEG data from different epileptic phases that we tested, the use of large number of time series resulted in that all methods except BTS perform poorly. For the MSCI data, again BTS performed best (even if marginally) in the majority of cases. Concluding,
our proposed BTS method for dynamic regression modeling turned out to have consistently
good prediction performance for all cases we studied.

\newpage

\bibliographystyle{elsarticle-harv}

\clearpage

\begin{table}
\caption{\label{tab:var2freqs} Most selected orders and the corresponding frequencies by
each method for the 4-d VAR(2) system and for different time series lengths.}
\begin{center}
\begin{tabular}{|l|l|l|l|}
\hline
$N$  & 100&200&400 \\
\hline
     & \multicolumn{3}{c|}{$y_{1,t+1}$ DR(2,0,0,2)}\\
\hline
FULL  & (2,0,0,2) 32\%  &(2,0,0,2) 69\%& (2,0,0,2) 90\% \\
\hline
VARB & (2,2,2,2) 97\%&(2,2,2,2) 100\%&(2,2,2,2) 100\%\\
\hline
CW & (0,0,1,1) 23\%&(0,0,1,1) 13\%& (5,1,1,1) 15\%\\
\hline
BTS & (2,0,0,1) 14\%&(2,0,0,2) 27\%&(2,0,0,2) 34\%\\
\hline\hline
     & \multicolumn{3}{c|}{$y_{2,t+1}$ DR(1,2,2,1)}\\
\hline
FULL  & (1,2,1,1) 50\%  &(1,2,1,1) 64\%& (1,2,1,1) 49\% \\
\hline
VARB & (2,2,2,2) 97\%&(2,2,2,2) 100\%&(2,2,2,2) 100\%\\
\hline
CW & (1,2,3,0) 27\%&(1,2,3,0) 29\%& (3,3,3,3) 20\%\\
\hline
BTS & (1,2,1,1) 35\%&(1,2,1,1,) 51\%&(1,2,1,1) 35\%\\
\hline\hline
     & \multicolumn{3}{c|}{$y_{3,t+1}$ DR(1,2,2,2)}\\
\hline
FULL  & (1,1,1,2) 26\%  &(1,2,1,2) 35\%& (1,2,2,2) 39\% \\
\hline
VARB & (2,2,2,2) 97\%&(2,2,2,2) 100\%&(2,2,2,2) 100\%\\
\hline
CW & (5,0,1,2) 16\%&(5,1,4,2) 20\%& (5,1,4,2) 44\%\\
\hline
BTS & (1,1,2,2) 32\%&(1,1,2,2) 40\%&(1,2,2,2) 43\%\\
\hline\hline
     & \multicolumn{3}{c|}{$y_{4,t+1}$ DR(1,1,0,2)}\\
\hline
FULL  & (1,1,0,2) 57\%  &(1,1,0,2) 86\%& (1,1,0,2) 94\% \\
\hline
VARB & (2,2,2,2) 97\%&(2,2,2,2) 100\%&(2,2,2,2) 100\%\\
\hline
CW & (0,0,1,2) 13\%&(5,0,3,2) 24\%& (5,0,4,2) 17\%\\
\hline
BTS & (1,1,0,2) 57\%&(1,1,0,2) 86\%&(1,1,0,2) 94\%\\
\hline
\end{tabular}
\end{center}
\end{table}

\clearpage

\begin{table}
\caption{\label{tab:var2nmse} Mean NMSE for 4-d VAR(2) model for each method, variable
and different time series lengths.}
\begin{center}\begin{tabular}{|l|c|c|c||c|c|c|}
\hline  $N$ &100 &200 &400 &100 &200 &400 \\
\hline
      & \multicolumn{3}{c||}{$y_{1,t+1}$} & \multicolumn{3}{c|}{$y_{2,t+1}$} \\
\hline $\mathrm{FULL_{ols}}$ &0.814 &0.711*&0.677*&0.333*&0.287*&0.274*\\\hline
$\mathrm{FULL_{pcr}}$ &0.816 &0.714*&0.677*&0.333*&0.288*&0.276*\\\hline
$\mathrm{FULL_{pls}}$ &0.815 &0.712*&0.677*&0.333*&0.287*&0.274*\\\hline
$\mathrm{FULL_{rr}}$  &0.805 &0.709*&0.677*&0.333*&0.287*&0.274*\\\hline
$\mathrm{VARB_{ols}}$ &0.796*&0.715*&0.684*&0.326*&0.287*&0.273*\\\hline
$\mathrm{VARB_{pcr}}$ &0.816 &0.719 &0.683*&0.330*&0.290*&0.274*\\\hline
$\mathrm{VARB_{pls}}$ &0.816 &0.718 &0.684*&0.328*&0.289*&0.273*\\\hline
$\mathrm{VARB_{rr}}$  &0.771*&0.709*&0.683*&0.325*&0.287*&0.273*\\\hline
$\mathrm{CW_{ols}}$   &0.920 &0.795 &0.714 &0.723 &0.557 &0.420 \\\hline
$\mathrm{CW_{pcr}}$   &0.925 &0.780 &0.716 &0.731 &0.562 &0.422 \\\hline
$\mathrm{CW_{pls}}$   &0.921 &0.798 &0.715 &0.730 &0.560 &0.421 \\\hline
$\mathrm{CW_{rr}}$    &0.908 &0.790 &0.713 &0.718 &0.559 &0.420 \\\hline
$\mathrm{MAX_{ols}}$  &0.996 &0.782 &0.714 &0.404 &0.315 &0.285 \\\hline
$\mathrm{MAX_{pcr}}$  &0.933 &0.783 &0.718 &0.379 &0.308 &0.282 \\\hline
$\mathrm{MAX_{pls}}$  &0.928 &0.779 &0.716 &0.377 &0.307 &0.282 \\\hline
$\mathrm{MAX_{rr}}$   &0.825 &0.736 &0.698 &0.384 &0.311 &0.284 \\\hline
$\mathrm{BTS_{ols}}$  &0.841 &0.724 &0.683*&0.359 &0.288*&0.274*\\\hline
$\mathrm{BTS_{pcr}}$  &0.845 &0.725 &0.683*&0.359 &0.289*&0.274*\\\hline
$\mathrm{BTS_{pls}}$  &0.843 &0.724 &0.683*&0.359 &0.288*&0.274*\\\hline
$\mathrm{BTS_{rr}}$   &0.832 &0.721 &0.683*&0.359 &0.288*&0.274*\\\hline
\end{tabular}
\end{center}
\end{table}

\clearpage

\begin{center}\begin{tabular}{|l|c|c|c||c|c|c|}
\hline  $N$ &100 &200 &400 &100 &200 &400 \\
\hline
      &   \multicolumn{3}{c||}{$y_{3,t+1}$} & \multicolumn{3}{c|}{$y_{4,t+1}$}\\
\hline  $N$  &100 &200 &400 &100 &200 &400\\
\hline $\mathrm{FULL_{ols}}$ &0.141 &0.104*&0.085*&0.213*&0.145*&0.116*\\\hline
$\mathrm{FULL_{pcr}}$ &0.141 &0.105*&0.086*&0.217*&0.148*&0.117*\\\hline
$\mathrm{FULL_{pls}}$ &0.141 &0.104*&0.085*&0.214*&0.145*&0.116*\\\hline
$\mathrm{FULL_{rr}}$  &0.141 &0.104*&0.085*&0.212*&0.145*&0.116*\\\hline
$\mathrm{VARB_{ols}}$ &0.135*&0.101*&0.084*&0.219*&0.148*&0.118*\\\hline
$\mathrm{VARB_{pcr}}$ &0.134*&0.102*&0.085*&0.221*&0.149*&0.118*\\\hline
$\mathrm{VARB_{pls}}$ &0.134*&0.101*&0.084*&0.220*&0.149*&0.118*\\\hline
$\mathrm{VARB_{rr}}$ &0.134*&0.101*&0.084*&0.218*&0.148*&0.117*\\\hline
$\mathrm{CW_{ols}}$   &0.198 &0.125 &0.090 &0.235 &0.158*&0.121
\\\hline $\mathrm{CW_{pcr}}$   &0.201 &0.125 &0.090 &0.236 &0.160 &0.122 \\\hline $\mathrm{CW_{pls}}$   &0.200 &0.125
&0.090 &0.234 &0.160 &0.122 \\\hline $\mathrm{CW_{rr}}$    &0.198 &0.125 &0.090 &0.236
&0.159 &0.121 \\\hline $\mathrm{MAX_{ols}}$  &0.166 &0.111 &0.088 &0.271 &0.161 &0.123
\\\hline $\mathrm{MAX_{pcr}}$  &0.157 &0.109 &0.088 &0.247 &0.160 &0.122 \\\hline
$\mathrm{MAX_{pls}}$  &0.156 &0.107 &0.087 &0.245 &0.160 &0.122 \\\hline
$\mathrm{MAX_{rr}}$   &0.155 &0.108 &0.087 &0.244 &0.158 &0.122 \\\hline
$\mathrm{BTS_{ols}}$ &0.139*&0.103*&0.085*&0.211*&0.145*&0.116*\\\hline
$\mathrm{BTS_{pcr}}$ &0.140*&0.104*&0.085*&0.216*&0.146*&0.116*\\\hline
$\mathrm{BTS_{pls}}$ &0.139*&0.103*&0.085*&0.212*&0.145*&0.116*\\\hline
$\mathrm{BTS_{rr}}$ &0.139*&0.103*&0.085*&0.211*&0.145*&0.116*\\\hline
\end{tabular}
\end{center}

\clearpage

\begin{table}
\caption{\label{tab:var2corerfreqs} As Table \ref{tab:var2freqs} but for the second 4-d
VAR(2) system.}
\begin{center}
\begin{tabular}{|l|l|l|l|}
\hline
$N$  & 100&200&400 \\
\hline
     & \multicolumn{3}{c|}{$y_{1,t+1}$ DR(2,0,0,2)}\\
\hline
FULL  & (2,0,0,2) 25\%  &(2,0,0,2) 60\%& (2,0,0,2) 88\% \\
\hline
VARB & (2,2,2,2) 98\%&(2,2,2,2) 100\%&(2,2,2,2) 100\%\\
\hline
CW & (0,0,1,1) 22\%&(4,0,1,1) 17\%& (5,0,1,1) 29\%\\
\hline
BTS & (2,0,1,0) 23\%&(2,0,1,0) 21\%&(2,0,1,2) 37\%\\
\hline\hline
     & \multicolumn{3}{c|}{$y_{2,t+1}$ DR(1,2,2,1)}\\
\hline
FULL  & (1,2,1,1) 41\%  &(1,2,1,1) 63\%& (1,2,1,1) 53\% \\
\hline
VARB & (2,2,2,2) 98\%&(2,2,2,2) 100\%&(2,2,2,2) 100\%\\
\hline
CW & (0,2,3,0) 1\%&(0,3,3,3) 1\%& (3,3,3,3) 51\%\\
\hline
BTS & (1,2,1,1) 25\%&(1,2,1,1,) 36\%&(1,2,1,1) 21\%\\
\hline\hline
     & \multicolumn{3}{c|}{$y_{3,t+1}$ DR(1,2,2,2)}\\
\hline
FULL  & (1,1,2,0) 35\%  &(1,2,1,2) 36\%& (1,1,2,2) 37\% \\
\hline
VARB & (2,2,2,2) 98\%&(2,2,2,2) 100\%&(2,2,2,2) 100\%\\
\hline
CW & (5,0,1,2) 30\%&(5,0,1,2) 28\%& (5,0,4,2) 34\%\\
\hline
BTS & (1,1,2,2) 22\%&(1,1,2,2) 33\%&(1,1,2,2) 31\%\\
\hline\hline
     & \multicolumn{3}{c|}{$y_{4,t+1}$ DR(1,1,0,2)}\\
\hline
FULL  & (1,1,0,2) 41\%  &(1,1,0,2) 79\%& (1,1,0,2) 93\% \\
\hline
VARB & (2,2,2,2) 98\%&(2,2,2,2) 100\%&(2,2,2,2) 100\%\\
\hline
CW & (5,0,1,2) 21\%&(5,0,2,2) 22\%& (5,0,3,2) 16\%\\
\hline
BTS & (1,1,0,2) 37\%&(1,1,0,2) 75\%&(1,1,0,2) 91\%\\
\hline
\end{tabular}
\end{center}
\end{table}

\clearpage

\begin{table}
\caption{\label{tab:var2corernmse} As Table \ref{tab:var2nmse} but for the second 4-d
 VAR(2).}
\begin{center}\begin{tabular}{|l|c|c|c||c|c|c|}
\hline  $N$ &100 &200 &400 &100 &200 &400 \\\hline
  & \multicolumn{3}{c||}{$y_{1,t+1}$} & \multicolumn{3}{c|}{$y_{2,t+1}$} \\\hline
$\mathrm{FULL_{ols}}$&0.830&0.727*&0.689*&0.430*&0.370*&0.354*\\\hline
$\mathrm{FULL_{pcr}}$&0.832&0.728*&0.690*&0.432*&0.370*&0.354*\\\hline
$\mathrm{FULL_{pls}}$&0.831&0.727*&0.690*&0.431*&0.370*&0.354*\\\hline
$\mathrm{FULL_{rr}}$&0.822&0.725*&0.689*&0.430*&0.370*&0.354*\\\hline
$\mathrm{VARB_{ols}}$&0.809*&0.729*&0.696*&0.415*&0.368*&0.353*\\\hline
$\mathrm{VARB_{pcr}}$&0.825&0.734&0.698*&0.415*&0.370*&0.353*\\\hline
$\mathrm{VARB_{pls}}$&0.822&0.733&0.697*&0.414*&0.368*&0.353*\\\hline
$\mathrm{VARB_{rr}}$&0.785*&0.722*&0.694*&0.414*&0.368*&0.353*\\\hline
$\mathrm{CW_{ols}}$&0.930&0.798&0.722&0.767&0.494&0.362\\\hline
$\mathrm{CW_{pcr}}$&0.931&0.807&0.725&0.770&0.498&0.364\\\hline
$\mathrm{CW_{pls}}$&0.930&0.802&0.724&0.768&0.495&0.361\\\hline
$\mathrm{CW_{rr}}$&0.919&0.794&0.721&0.760&0.494&0.362\\\hline
$\mathrm{MAX_{ols}}$&1.012&0.798&0.726&0.509&0.404&0.368\\\hline
$\mathrm{MAX_{pcr}}$&0.953&0.792&0.726&0.487&0.397&0.366\\\hline
$\mathrm{MAX_{pls}}$&0.944&0.790&0.725&0.482&0.394&0.364\\\hline
$\mathrm{MAX_{rr}}$&0.841&0.750&0.709&0.486&0.400&0.368\\\hline
$\mathrm{BTS_{ols}}$&0.857&0.737&0.699*&0.465&0.370*&0.355*\\\hline
$\mathrm{BTS_{pcr}}$&0.858&0.737&0.699*&0.467&0.370*&0.356*\\\hline
$\mathrm{BTS_{pls}}$&0.858&0.737&0.699*&0.466&0.370*&0.356*\\\hline
$\mathrm{BTS_{rr}}$&0.848&0.734&0.699*&0.465&0.370*&0.355*\\\hline
\end{tabular}
\end{center}
\end{table}

\clearpage

\begin{center}\begin{tabular}{|l|c|c|c||c|c|c|}
\hline  $N$ &100 &200 &400 &100 &200 &400 \\\hline
    &  \multicolumn{3}{c||}{$y_{3,t+1}$} & \multicolumn{3}{c|}{$y_{4,t+1}$}\\\hline
$\mathrm{FULL_{ols}}$&0.160&0.119*&0.092*&0.220*&0.157*&0.119*\\\hline
$\mathrm{FULL_{pcr}}$&0.160&0.120&0.092*&0.221*&0.157*&0.119*\\\hline
$\mathrm{FULL_{pls}}$&0.160&0.119*&0.092*&0.221*&0.157*&0.119*\\\hline
$\mathrm{FULL_{rr}}$&0.159*&0.119*&0.092*&0.220*&0.157*&0.119*\\\hline
$\mathrm{VARB_{ols}}$&0.154*&0.115*&0.091*&0.222*&0.158*&0.120*\\\hline
$\mathrm{VARB_{pcr}}$&0.156*&0.116*&0.091*&0.224*&0.158*&0.120*\\\hline
$\mathrm{VARB_{pls}}$&0.155*&0.116*&0.091*&0.223*&0.158*&0.120*\\\hline
$\mathrm{VARB_{rr}}$&0.154*&0.115*&0.091*&0.221*&0.158*&0.120*\\\hline
$\mathrm{CW_{ols}}$&0.203&0.144&0.107&0.237&0.166&0.124\\\hline
$\mathrm{CW_{pcr}}$&0.207&0.147&0.108&0.237&0.167&0.124\\\hline
$\mathrm{CW_{pls}}$&0.206&0.145&0.107&0.237&0.167&0.124\\\hline
$\mathrm{CW_{rr}}$&0.203&0.144&0.107&0.237&0.166&0.124\\\hline
$\mathrm{MAX_{ols}}$&0.190&0.126&0.095&0.276&0.173&0.126\\\hline
$\mathrm{MAX_{pcr}}$&0.182&0.125&0.095&0.258&0.170&0.124\\\hline
$\mathrm{MAX_{pls}}$&0.180&0.124&0.095&0.257&0.169&0.124\\\hline
$\mathrm{MAX_{rr}}$&0.177&0.123&0.094&0.248&0.169&0.125\\\hline
$\mathrm{BTS_{ols}}$&0.162&0.118*&0.092*&0.223*&0.157*&0.119*\\\hline
$\mathrm{BTS_{pcr}}$&0.163&0.121&0.092*&0.224*&0.157*&0.119*\\\hline
$\mathrm{BTS_{pls}}$&0.162&0.118*&0.092*&0.223*&0.157*&0.119*\\\hline
$\mathrm{BTS_{rr}}$&0.161&0.118*&0.092*&0.222*&0.157*&0.119*\\\hline
\end{tabular}
\end{center}

\clearpage

\begin{table}
\begin{center}
\caption{\label{tab:Sj}Efficiency scores for the 81 bivariate DR systems.}
\begin{tabular}{|l|c|c|c|c|c|c|c|c|c|c|c|c|c|c|c|c|c|c|c|c|} \hline
 &$\mathrm{FULL_{ols}}$ &$\mathrm{FULL_{pcr}}$ &$\mathrm{FULL_{pls}}$ &$\mathrm{FULL_{rr}}$ \\\hline
$N$=100 &1.340 &1.370 &1.361 &1.344 \\\hline $N$=200 &0.213 &0.220 &0.217 &0.216 \\\hline
$N$=400 &0.041 &0.042 &0.042 &0.041 \\\hline\hline
 &$\mathrm{VARB_{ols}}$ &$\mathrm{VARB_{pcr}}$ &$\mathrm{VARB_{pls}}$ &$\mathrm{VARB_{rr}}$\\\hline
$N$=100 &1.240 &1.311 &1.306 &1.211 \\\hline $N$=200 &0.241 &0.257 &0.255 &0.237 \\\hline
$N$=400 &0.050 &0.054 &0.053 &0.049 \\\hline\hline
 &$\mathrm{CW_{ols}}$ &$\mathrm{CW_{pcr}}$ &$\mathrm{CW_{pls}}$ &$\mathrm{CW_{rr}}$\\\hline
$N$=100 &5.961 &6.103 &6.051 &6.011 \\\hline $N$=200 &1.868 &1.902 &1.890 &1.874 \\\hline
$N$=400 &0.505 &0.511 &0.510 &0.505 \\\hline\hline
&$\mathrm{MAX_{ols}}$ &$\mathrm{MAX_{pcr}}$ &$\mathrm{MAX_{pls}}$
&$\mathrm{MAX_{rr}}$\\\hline $N$=100 &2.432 &2.383 &2.370 &1.772 \\\hline $N$=200 &0.463
&0.481 &0.478 &0.382 \\\hline $N$=400 &0.099 &0.108 &0.105 &0.088 \\\hline\hline
&$\mathrm{BTS_{ols}}$ &$\mathrm{BTS_{pcr}}$ &$\mathrm{BTS_{pls}}$
&$\mathrm{BTS_{rr}}$\\\hline $N$=100 &1.890 &1.923 &1.919 &1.902 \\\hline $N$=200 &0.248
&0.253 &0.253 &0.251 \\\hline $N$=400 &0.045 &0.046 &0.046 &0.045 \\\hline
\end{tabular}
\end{center}
\end{table}

\clearpage

\begin{table*}\begin{center}
\caption{\label{tab:C8} Average NMSE for $x'_{8,t}$ of the first system of 8
multi-collinear time series for selected methods as given in the first column and
different time series lengths $N$ and strength of collinearity c.}
\begin{tabular}{|l|c|c|c||c|c|c|}
\hline $N$ &100 &200 &400 &100 &200 &400 \\
\hline
      & \multicolumn{3}{c||}{c=0} & \multicolumn{3}{c|}{c=0.5} \\
\hline $\mathrm{FULL_{ols}}$ &0.694*&0.615*&0.587*&0.617*&0.539*&0.519*\\\hline
$\mathrm{VARB_{ols}}$ &0.741 &0.635 &0.597 &0.641*&0.553*&0.527 \\\hline
$\mathrm{CW_{ols}}$   &0.825 &0.637 &0.590*&0.673 &0.560*&0.534 \\\hline
$\mathrm{MAX_{ols}}$  &1.001 &0.717 &0.634 &0.860 &0.626 &0.558 \\\hline
$\mathrm{MAX_{rr}}$   &0.825 &0.679 &0.621 &0.734 &0.599 &0.549 \\\hline
$\mathrm{BTS_{ols}}$  &0.718*&0.616*&0.587*&0.643*&0.542*&0.519*\\\hline
\hline
      &  \multicolumn{3}{c||}{c=1} & \multicolumn{3}{c|}{c=2}\\\hline
$\mathrm{FULL_{ols}}$&0.592*&0.506*&0.480*&0.569*&0.501*&0.469*\\\hline
$\mathrm{VARB_{ols}}$&0.598*&0.514*&0.487*&0.580*&0.498*&0.469*\\\hline
$\mathrm{CW_{ols}}$  &0.604*&0.526 &0.494 &0.580*&0.504*&0.473*\\\hline
$\mathrm{MAX_{ols}}$ &0.795 &0.585 &0.516 &0.777 &0.569 &0.497 \\\hline
$\mathrm{MAX_{rr}}$  &0.655 &0.556 &0.508 &0.585*&0.516 &0.484 \\\hline
$\mathrm{BTS_{ols}}$ &0.602*&0.514*&0.481*&0.569*&0.503*&0.471*\\\hline
\end{tabular}
\end{center}
\end{table*}

\clearpage

\begin{table*}
\caption{\label{tab:C82}As Table \ref{tab:C8} but for $x'_{8,t}$ of the second system of
8 multi-collinear time series.}
\begin{center}\begin{tabular}{|l|c|c|c||c|c|c|}
\hline $N$  &100 &200 &400 &100 &200 &400\\
\hline
      & \multicolumn{3}{c||}{c=0} & \multicolumn{3}{c|}{c=0.5} \\
\hline $\mathrm{FULL_{ols}}$ &0.701*&0.617*&0.588*&0.691*&0.580*&0.555*\\\hline
$\mathrm{VARB_{ols}}$ &0.780 &0.638*&0.598*&0.814 &0.717 &0.686 \\\hline
$\mathrm{VARB_{rr}}$  &0.770 &0.635 &0.597*&0.807 &0.714 &0.684 \\\hline
$\mathrm{CW_{ols}}$   &0.827 &0.634*&0.592*&0.844 &0.697 &0.613 \\\hline
$\mathrm{CW_{rr}}$    &0.828 &0.633*&0.591*&0.842 &0.697 &0.614 \\\hline
$\mathrm{MAX_{ols}}$  &1.001 &0.723 &0.635 &0.872 &0.666 &0.593 \\\hline
$\mathrm{MAX_{rr}}$   &0.833 &0.684 &0.622 &0.756 &0.638 &0.585 \\\hline
$\mathrm{BTS_{ols}}$ &0.719*&0.618*&0.588*&0.749 &0.584*&0.556*\\\hline
\hline
      &  \multicolumn{3}{c||}{c=1} & \multicolumn{3}{c|}{c=2}\\\hline
 $\mathrm{FULL_{ols}}$ &0.654*&0.560*&0.535*&0.622*&0.563*&0.532*\\\hline $\mathrm{VARB_{ols}}$ &0.848 &0.773 &0.747
&0.870 &0.802 &0.780 \\\hline $\mathrm{VARB_{rr}}$  &0.838 &0.771 &0.745 &0.837 &0.797
&0.777 \\\hline $\mathrm{CW_{ols}}$   &0.681*&0.597 &0.562 &0.708 &0.598 &0.552 \\\hline
$\mathrm{CW_{rr}}$    &0.674*&0.593 &0.558 &0.625*&0.572*&0.545 \\\hline
$\mathrm{MAX_{ols}}$  &0.842 &0.636 &0.570 &0.826 &0.627 &0.561 \\\hline
$\mathrm{MAX_{rr}}$   &0.702 &0.609 &0.563 &0.633*&0.577*&0.548 \\\hline
$\mathrm{BTS_{ols}}$ &0.676*&0.572*&0.538*&0.625*&0.570*&0.538*\\\hline
\end{tabular}
\end{center}
\end{table*}

\clearpage

\begin{table}\begin{center}
\caption{\label{tab:AverMedianNmseInc} Averaged percent change of median NMSE over all
channels for the 3 records, along with its standard deviation in parenthesis. The
reference method is BTS with OLS estimation of parameters.}

\begin{tabular}{|c|c|c|c|}
\hline 1st record    & ols           & pls          & rr           \\
\hline VARB& 15.2 (6.4) \% & 9.4 (6.7) \% & 18.3 (9.6) \%\\
\hline CW  & 10.0 (3.8) \% & 6.5 (4.6) \% & 13.7 (7.5) \%\\
\hline MAX & 14.3 (2.9) \% & 8.0 (4.3) \% & 15.7 (7.6) \%\\
\hline BTS &   0        \% &-0.2 (0.5) \% &  0.7 (0.9) \%\\
\hline 2nd record    &            &          &           \\
\hline VARB& 12.4 (3.1) \% & 7.0 (4.2) \% & 11.0 (5.2) \%\\
\hline CW  &  9.2 (1.9) \% & 5.5 (3.2) \% &  8.6 (3.9) \%\\
\hline MAX & 11.3 (1.9) \% & 5.6 (2.7) \% &  8.2 (3.7) \%\\
\hline BTS &   0        \% &-0.1 (0.3) \% &  0.4 (0.7) \%\\
\hline 3rd record    &            &          &            \\
\hline VARB& 51.3 (10.6) \% &32.9 (9.9) \% & 34.6 (9.6) \%\\
\hline CW  & 35.9 (5.3) \% & 22.5 (7.3) \% & 22.8 (7.6) \%\\
\hline MAX & 37.9 (6.7) \% & 22.4 (6.9) \% & 23.3 (7.6) \%\\
\hline BTS &   0        \% &-0.2 (1.0) \% &  0.3 (0.7) \%\\
\hline
\end{tabular}
\end{center}
\end{table}

\clearpage

\begin{table}\begin{center}
\caption{\label{tab:AverAUC} Average AUC value over all channels.}
\begin{tabular}{|c|c|c|c|}
\hline 1st-2nd records&ols&pls&rr\\
\hline VARB&0.626&0.634&0.612\\
\hline CW  &0.634&0.637&0.619\\
\hline MAX &0.627&0.634&0.612\\
\hline BTS &0.637&0.638&0.636\\
\hline 2nd-3rd records&&&    \\
\hline VARB&0.757&0.784&0.779\\
\hline CW  &0.766&0.792&0.790\\
\hline MAX &0.756&0.786&0.784\\
\hline BTS &0.789&0.789&0.790\\
\hline
\end{tabular}
\end{center}
\end{table}

\clearpage

\begin{table}\begin{center}
\caption{\label{tab:MSCIDEV} Average NMSE of the data of MSC index for the set of developed countries across the four split samples. Selected methods are shown, and the last row of the table shows the average across all countries.}
\begin{tabular}{|c|l|c|c|c|c|c|c|c|}
\hline
&&\textbf{$\mathrm{VARB_{ols}}$}&\textbf{$\mathrm{CW_{ols}}$}&\textbf{$\mathrm{CW_{rr}}$}&\textbf{$\mathrm{MAX_{ols}}$}&\textbf{$\mathrm{MAX_{rr}}$}&\textbf{$\mathrm{BTS_{ols}}$}&\textbf{$\mathrm{BTS_{rr}}$}\\\hline
1&Australia&1.001&0.723&0.702&0.913&0.723&0.679&0.681\\\hline
2&New Zealand&1.004&0.848&0.828&1.194&0.863&0.807&0.808\\\hline
3&Japan&1.001&0.829&0.764&1.100&0.791&0.770&0.767\\\hline
4&Hong Kong&1.001&0.900&0.858&1.230&0.865&0.875&0.847\\\hline
5&Singapore&1.001&0.930&0.883&1.126&0.883&0.884&0.880\\\hline
6&Austria&1.002&0.918&0.914&1.111&0.922&0.895&0.896\\\hline
7&Belgium&1.002&0.946&0.944&1.240&0.969&0.920&0.919\\\hline
8&Denmark&1.001&0.925&0.923&1.182&0.942&0.930&0.926\\\hline
9&Finland&1.003&0.961&0.943&1.303&0.957&0.960&0.954\\\hline
10&France&1.001&0.885&0.888&1.133&0.928&0.900&0.898\\\hline
11&Germany&1.001&0.960&0.957&1.205&0.955&0.922&0.919\\\hline
12&Greece&1.002&0.941&0.933&1.279&0.974&0.941&0.941\\\hline
13&Ireland&1.002&0.923&0.921&1.179&0.964&0.917&0.915\\\hline
14&Italy&1.002&0.952&0.952&1.139&0.978&0.932&0.932\\\hline
15&Netherlands&1.001&0.949&0.944&1.204&0.947&0.901&0.900\\\hline
16&Norway&1.001&0.924&0.902&1.066&0.914&0.911&0.904\\\hline
17&Portugal&1.002&0.975&0.959&1.153&0.954&0.962&0.955\\\hline
18&Spain&1.001&0.986&0.970&1.226&0.976&0.960&0.947\\\hline
19&Sweden&1.001&0.916&0.914&1.201&0.940&0.923&0.918\\\hline
20&Switzerland&1.001&0.908&0.912&1.243&0.962&0.923&0.920\\\hline
21&UK&1.001&0.893&0.895&1.119&0.923&0.854&0.856\\\hline
22&Canada&1.002&0.993&0.993&1.313&1.014&1.022&1.015\\\hline
23&USA&1.001&1.018&1.011&1.408&1.012&1.011&1.008\\\hline
&\bf{MEAN}&1.002&0.922&0.909&1.186&0.929&0.904&0.900\\\hline
\end{tabular}
\end{center}
\end{table}

\clearpage

\centerline{\hbox{\includegraphics[height=8cm]{Fig1a}
\includegraphics[height=8cm]{Fig1b}}}
\centerline{\hbox{\includegraphics[height=8cm]{Fig1c}
\includegraphics[height=8cm]{Fig1d}}}

\centerline{Figure 1}

\clearpage

 \centerline{\includegraphics[width=17cm]{Fig2.eps}}

\centerline{Figure 2}

\clearpage

\centerline{\hbox{\includegraphics[height=5cm]{Fig3a.eps}
\includegraphics[height=5cm]{Fig3b.eps}}}
\centerline{\hbox{\includegraphics[height=5cm]{Fig3c.eps}
\includegraphics[height=5cm]{Fig3d.eps}}}

\centerline{Figure 3}

\end{document}